\documentclass[runningheads]{llncs}
\usepackage[T1]{fontenc}
\usepackage{graphicx}
\usepackage{amsfonts}
\usepackage{amsmath}
\usepackage[table]{xcolor}
\usepackage{tikz}
\usepackage{multirow}
\usepackage{algorithm}
\usepackage{siunitx} 
\usepackage{algcompatible}
\usepackage{algpseudocode}
\usepackage{orcidlink}

\usepackage{hyperref}

\begin{document}
\title{cWDM: Conditional Wavelet Diffusion Models for Cross-Modality 3D Medical Image Synthesis}
\titlerunning{cWDM for Cross-Modality 3D Medical Image Synthesis}
%
\author{Paul Friedrich\orcidlink{0000-0003-3653-5624} \and 
        Alicia Durrer\orcidlink{0009-0007-8970-909X} \and 
        Julia Wolleb\orcidlink{0000-0003-4087-5920} \and 
        Philippe C. Cattin\orcidlink{0000-0001-8785-2713}}
\authorrunning{P. Friedrich et al.}

\institute{Department of Biomedical Engineering, University of Basel, Allschwil, Switzerland\\
\email{paul.friedrich@unibas.ch}}
\maketitle              
\begin{abstract}
This paper contributes to the "BraTS 2024 Brain MR Image Synthesis Challenge" and presents a conditional Wavelet Diffusion Model (cWDM) for directly solving a paired image-to-image translation task on high-resolution volumes. While deep learning-based brain tumor segmentation models have demonstrated clear clinical utility, they typically require MR scans from various modalities (T1, T1ce, T2, FLAIR) as input. However, due to time constraints or imaging artifacts, some of these modalities may be missing, hindering the application of well-performing segmentation algorithms in clinical routine. To address this issue, we propose a method that synthesizes one missing modality image conditioned on three available images, enabling the application of downstream segmentation models. We treat this paired image-to-image translation task as a conditional generation problem and solve it by combining a Wavelet Diffusion Model for high-resolution 3D image synthesis with a simple conditioning strategy. This approach allows us to directly apply our model to full-resolution volumes, avoiding artifacts caused by slice- or patch-wise data processing. While this work focuses on a specific application, the presented method can be applied to all kinds of paired image-to-image translation problems, such as CT~$\leftrightarrow$~MR and MR~$\leftrightarrow$~PET translation, or mask-conditioned anatomically guided image generation.
\end{abstract}
\section{Introduction}
\begin{figure}
    \centering
    \includegraphics[width=\textwidth]{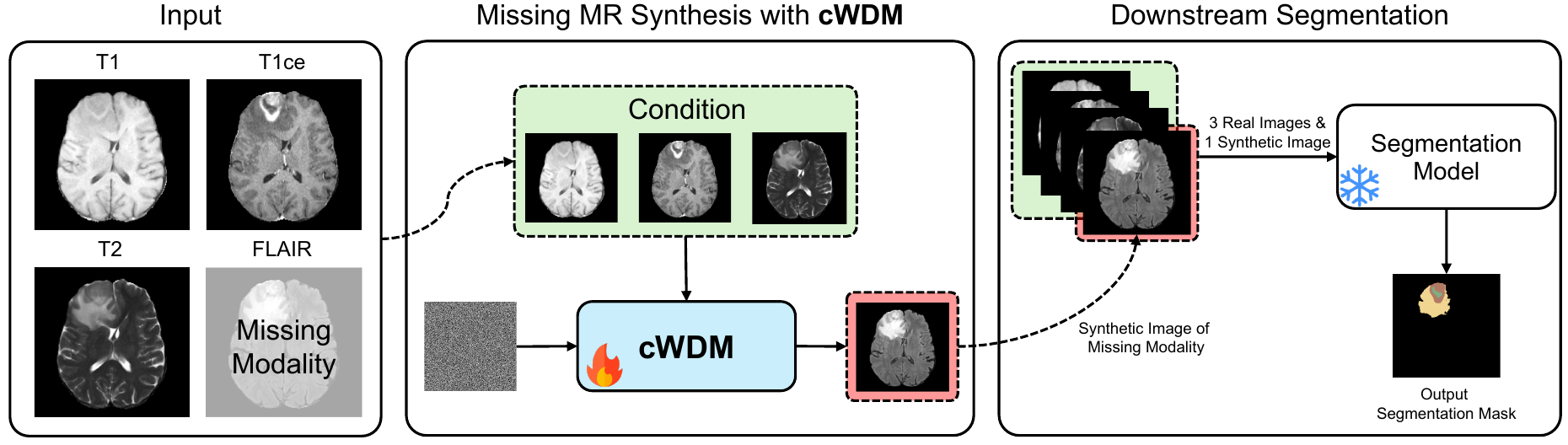}
    \caption{Schematic overview of the proposed pipeline for missing MR image generation - in this case, for a missing FLAIR image. We aim to generate the missing modality image conditioned on the three available images, ultimately allowing for pre-trained downstream task segmentation models to be applied. The same principle applies if another imaging modality is missing. For simplicity, all 3D volumes are displayed as 2D slices.}
    \label{fig:overview}
\end{figure}

Deep learning-based brain tumor segmentation methods have proven to be valuable tools for automating manual segmentation tasks in magnetic resonance images, supporting physicians, and achieving clear clinical utility \cite{li2023brain}. Most of these methods require four different MR images of different modalities as input: T1-weighted images with (T1ce) and without (T1) contrast enhancement, T2-weighted images (T2), as well as T2-weighted fluid attenuated inversion recovery (FLAIR) images. Time constraints or imaging artifacts, e.g. due to patient motion, can lead to the problem of missing MR sequences, hindering the application of automatic segmentation algorithms in the clinical routine. This paper contributes to the "BraTS 2024 Brain MR Image Synthesis Challenge" and tries to solve this problem by applying deep generative models to synthesize missing modality MR images, enabling the application of automatic segmentation methods. The general setting is shown in Fig. \ref{fig:overview}.
Due to the three-dimensional nature of medical images and the computational cost of modeling these volumes, most existing methods for generating 3D medical images operate on patches or slices of the volumes, stitching them together after processing them separately. This usually results in inter-slice or inter-patch inconsistencies. Following the recent success of Wavelet Diffusion Models (WDMs) in generating high-resolution medical images \cite{friedrich2024wdm}, we explore and extend this framework proposing \textbf{cWDM}, a \textbf{c}onditional \textbf{W}avelet \textbf{D}iffusion \textbf{M}odel that allows for solving paired image-to-image translation tasks on high-resolution medical volumes.
\subsubsection{Related Work} Several GAN-based approaches, such as pix2pix \cite{isola2017image}, solve image-to-image translation tasks by modeling the conditional distribution of target domain images given paired source domain images. Other GAN-based approaches apply more sophisticated methods, such as solving unpaired translation via cycle consistency \cite{zhu2017unpaired}, or using disentangled representations \cite{lee2018diverse} to solve the image-to-image translation task. Although GAN-based methods have widely been used for 3D medical image-to-image translation \cite{hu2021bidirectional,kalantar2023non,uzunova2020memory,zhao2021mri} and offer several advantages, including fast sampling speed, they are difficult to train, especially on 3D data. Furthermore, they are prone to problems such as mode collapse.
Denoising Diffusion Models \cite{ho2020denoising,sohl2015deep} have outperformed GANs on image synthesis \cite{dhariwal2021diffusion} and have widely been applied to solve image-to-image translation tasks \cite{saharia2022palette}. They have already been adapted for translating 3D medical volumes \cite{durrer2023diffusion,friedrich2023point,graf2023denoising,kim024aldm,pan2023cycle,zhu2023make} and have shown promising results. Due to the computational complexity of Denoising Diffusion Models, these approaches primarily operate on 2D slices \cite{durrer2023diffusion}, apply pseudo-3D approaches \cite{zhu2023make} or operate on learned compressed latent representations of the data \cite{kim024aldm}, which can be hard to obtain from high-resolution 3D medical images \cite{friedrich2024deep}. In this work, we apply Wavelet Diffusion Models \cite{friedrich2024wdm,guth2022wavelet,phung2023wavelet} to efficiently solve the paired image-to-image translation task on full-resolution 3D volumes. 
\section{Background}
Wavelet Diffusion Models \cite{friedrich2024wdm} are a type of Denoising Diffusion Model \cite{dhariwal2021diffusion,ho2020denoising} that operate on wavelet decomposed images $x = \text{DWT}(y)$, with $x \in \mathbb{R}^{8 \times \frac{D}{2} \times \frac{H}{2} \times \frac{W}{2}}$
rather than the original images $y \in \mathbb{R}^{D \times H \times W}$ themselves. Their general concept is similar to that of Latent Diffusion Models \cite{rombach2022high}, but they replace the first-stage autoencoder with a training-free approach for spatial dimensionality reduction, namely the discrete wavelet transform (DWT). The final output images of WDMs are produced by generating synthetic wavelet coefficients $\tilde{x}_0$ and applying Inverse DWT (IDWT) to them.

Following \cite{ho2020denoising,nichol2021improved}, we define a \emph{forward diffusion process} that gradually perturbs a sample $x$, in this case the wavelet coefficients of an image, with Gaussian noise. This noise perturbation follows a sequence of normal distributions with a pre-defined variance schedule $\beta_{1:T}$ and a given number of timesteps $T$, with $t\in\{1,...,T\}$:
\begin{equation}
    q(x_{t}|x_{t-1}):=\mathcal{N}(\sqrt{1-\beta_t}x_{t-1},\beta_t\boldsymbol{I}).
\end{equation}
For large $T$, this forward diffusion process maps a sample $x_0$ to a standard normal distribution $x_T\sim\mathcal{N}(0,\boldsymbol{I})$. The \emph{learned reverse process} aims to run this forward noising process backward in time to generate new samples by drawing $x_T\sim\mathcal{N}(0,\boldsymbol{I})$ and passing it through this reverse process. We can model such a reverse process as a Markov chain
\begin{equation}
    p_{\theta}(x_{0:T}):=p(x_T)\prod_{t=1}^{T}p_{\theta}(x_{t-1}|x_t, \tilde{x}_0),
\end{equation}
with each transition following a normal distribution with mean $\mu_t(x_t, \tilde{x}_0)$ parameterized by a time-conditioned neural network $\epsilon_\theta$. This neural network is trained to predict the denoised wavelet coefficients $\tilde{x}_0=\epsilon_{\theta}(x_t,t)$ using a Mean Squared Error (MSE) loss between predicted and ground truth wavelet coefficients:
\begin{equation}
    \mathcal{L}_{MSE} = \|\tilde{x}_0 - x_0\|^{2}_{2}.
\end{equation}
Each transition of the reverse process can then be described in the following form:
\begin{equation}
    p_{\theta}(x_{t-1}|x_t, \tilde{x}_0):=\mathcal{N}(\mu_t(x_t, \tilde{x}_0), \tilde{\beta}_t\boldsymbol{I}),
\end{equation}
with its learned mean $\mu_t$ and pre-defined variance $\tilde{\beta}_t $ being formulated as:
\begin{equation}
   \mu_t(x_t, \tilde{x}_0) := \frac{\sqrt{\bar{\alpha}_t-1}\beta_t}{1-\bar{\alpha}_t}\tilde{x}_0 + \frac{\sqrt{\alpha_t}(1-\bar{\alpha}_{t-1})}{1-\bar{\alpha}_t}x_t,
\end{equation}
\begin{equation}
    \tilde{\beta}_t := \frac{1-\bar{\alpha}_{t-1}}{1-\bar{\alpha}_t}\beta_t,
\end{equation}
with $\alpha_t:=1-\beta_t$ and $\bar{\alpha}_t := \prod_{s=1}^t\alpha_s$. A detailed description for training and sampling from these models can be found in \cite{friedrich2024wdm}.

\section{Method}
\begin{figure}
    \centering
    \includegraphics[width=\textwidth]{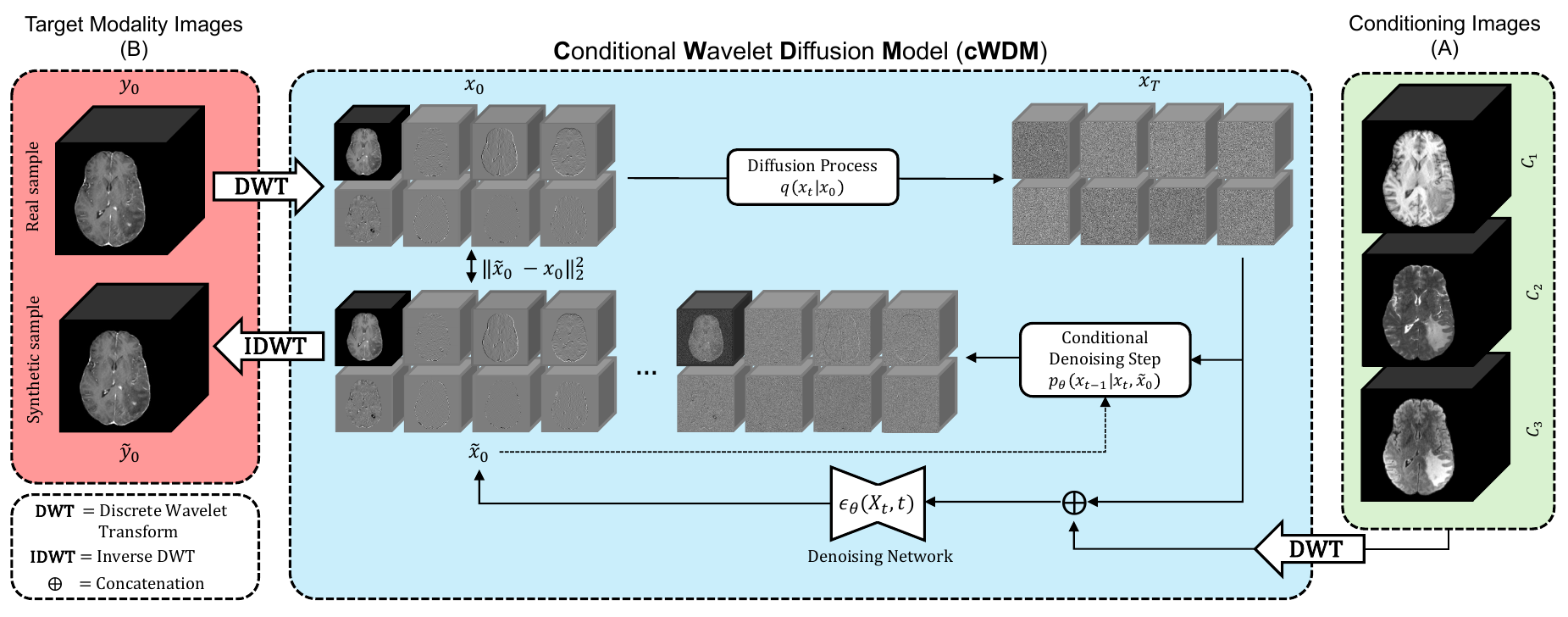}
    \caption{Schematic overview of the proposed conditional Wavelet Diffusion Model - in this case for a missing T1ce image. The process of generating the wavelet coefficients $\tilde{x}_0$ of the output images is conditioned on the wavelet coefficients of the conditioning images by concatenating them with the noisy coefficients in each denoising step.}
    \label{fig:cwdm}
\end{figure}

We define cross-modality image synthesis as an image-to-image translation task, where we aim to find a mapping function $F: A \rightarrow B$ that maps from a source domain $A$ to a target domain $B$. In a setting where paired data of the form $\{a_i,b_i\}_{i=1}^N$, with $a_i \in A$ and $b_i \in B$ is available, this problem can be formulated as a conditional generation task with the mapping function $F$ being a deep generative model that models the conditional distribution $p(b|a)$. In our case, we apply a 3D wavelet diffusion model \cite{friedrich2024wdm} with a Palette-like conditioning strategy \cite{saharia2022palette} to model this conditional distribution by conditioning the generation process of a target modality image on multiple source domain images. We do this by concatenating the conditioning image's wavelet coefficients in each denoising step of the diffusion model. This is shown in Fig.~\ref{fig:cwdm}. 
Our model $\epsilon_{\theta}(X_t, t)$ therefore learns to generate the denoised wavelet coefficients $\tilde{x}_0$ of the target domain image given a noisy version $x_t$ at timestep $t$, as well as the condition $c = \text{DWT}(C_1)\oplus \text{DWT}(C_2) \oplus \text{DWT}(C_3)$. The input to the model is then defined by concatenating the conditioning wavelet coefficients $c\in\mathbb{R}^{24 \times \frac{D}{2} \times \frac{H}{2} \times \frac{W}{2}}$ in the channel dimension $X_t = x_t \oplus c$, such that $X_t\in\mathbb{R}^{32 \times \frac{D}{2} \times \frac{H}{2} \times \frac{W}{2}}$. To allow for the generation of multiple missing contrasts, we trained four separate models, where each of the models learns to generate images of one modality, conditioned on three given images from the other modalities. When processing a case, we first detect the missing modality, choose the correct model to run inference and follow the conditional sampling strategy described in Algorithm~\ref{alg:Sampling}. In an exemplary case of a missing FLAIR image, $C_1$ would be the T1-weighted image, $C_2$ would be the contrast-enhanced T1-weighted image, and $C_3$ would be the T2-weighted image. The model would output a synthetic FLAIR image.
\begin{algorithm}[H]
    \caption{Conditional Sampling}\label{alg:Sampling}
    \begin{algorithmic}
        \State\vphantom{$\|_{a}^{b}$}\textbf{Input:} Conditioning Images $C_1$, $C_2$, $C_3$
        \State\vphantom{$\|_{a}^{b}$}\textbf{Output:} Missing Modality Image $\tilde{y}_0$\\
        \State\vphantom{$\|_{a}^{b}$}$x_T \sim \mathcal{N}(0,\boldsymbol{I})$
        \State\vphantom{$\|_{a}^{b}$}$c = \text{DWT}(C_1)\oplus \text{DWT}(C_2) \oplus \text{DWT}(C_3)$
        \For{$t =T, ..., 1$}\vphantom{$\|_{a}^{b}$}
        \State\vphantom{$\|_{a}^{b}$}$X_t = x_t \oplus c$
        \State\vphantom{$\|_{a}^{b}$}$\tilde{x}_0 = \epsilon_\theta(X_t, t)$
        \State\vphantom{$\|_{a}^{b}$}$x_{t-1} \sim p_\theta(x_{t-1}|x_t, \tilde{x}_0)$
        \EndFor\vphantom{$\|_{a}^{b}$}
        \State\vphantom{$\|_{a}^{b}$}$\tilde{y}_0 = \text{IDWT}(x_0)$
        \State\vphantom{$\|_{a}^{b}$}\textbf{return} $\tilde{y}_0$
    \end{algorithmic}
\end{algorithm}

\section{Experiments}
\subsection{Experimental Settings}
\subsubsection{Dataset}
We evaluate our model's performance on the dataset provided by the challenge organizers. A detailed description of the dataset, which is based on the RSNA-ASNR-MICCAI BraTS 2021 dataset \cite{bakas2017advancing,bakas2018identifying,karargyris2023federated,menze2014multimodal} can be found in \cite{li2023brain}. The dataset contains a collection of multi-parametric MRI (mpMRI) scans of brain tumors from various institutions, as well as segmentation masks for different tumor sub-regions. The provided data contains 1251 training and 219 validation cases, each providing T1, T1ce, T2 and FLAIR images with a resolution of $155 \times 240 \times 240$. The training data additionally contains the ground truth segmentation masks. We preprocessed all volumes by clipping the upper and lower 0.1 percentile intensity values and by normalizing them to a range of [0,1].
\subsubsection{Training}
We trained a total of four models to solve the task by selecting one of the modalities in the training set as the target and the other three as the condition. Thus, each of these four models learned to generate one of the four modalities conditioned on three images of the other modalities. We want to note that training a single model for solving this task would also be possible and is arguably the more efficient and scalable solution. However, as we expected a slightly worse performance than the multi-model approach, we decided to train separate models. All models were trained for \SI{1.2}{M} iterations, using an Adam optimizer with a learning rate of $1 \times 10^{-5}$ and a batch size of $1$. All experiments were carried out on a single NVIDIA A100 (\SI{40}{\giga\byte}) GPU.
\subsubsection{Implementation Details} 
We define a diffusion process with $T=1000$ timesteps and a linear variance schedule between $\beta_1 = 1 \times 10^{-4}$ and $\beta_T=0.02$. For the denoising model $\epsilon_{\theta}$, we follow an implementation from \cite{friedrich2024wdm}, change the skip connections from additive \cite{bieder2023memory} to standard ones, and set the number of base channels to $C=64$. An ablation study for the choice of these hyperparameters is provided in Section~\ref{subsec:ablation}. The source code is publicly available at \url{https://github.com/pfriedri/cwdm}.
\subsubsection{Evaluation Metrics}
We evaluate the quality of the generated missing modality images using Mean Squared Error (MSE), Peak Signal-to-Noise-Ratio (PSNR), as well as Structural Similarity Index Measure (SSIM). The scores are computed on the complete, normalized volumes of the validation dataset.
\subsection{Results on Validation Data}
In Tab.~\ref{tab:brats}, we report quantitative evaluation scores for all four models by dropping the modality that the corresponding model was trained to generate. In addition, we report scores on a pseudo-validation set by randomly dropping one modality for each subject. We create this pseudo-validation set using a script provided by the challenge organizers. It was not possible to report downstream segmentation results for the validation set, as the challenge organizers did not provide a way to obtain these scores, nor did they provide ground truth segmentation masks.

\begin{figure}
    \centering
    \resizebox{\textwidth}{!}{
        \begin{tikzpicture}
            \fill (0.1, 0.1) rectangle (8.125, 12.125);
            \node[] at (0, 2)[anchor=south west]  {\includegraphics[width=2cm]{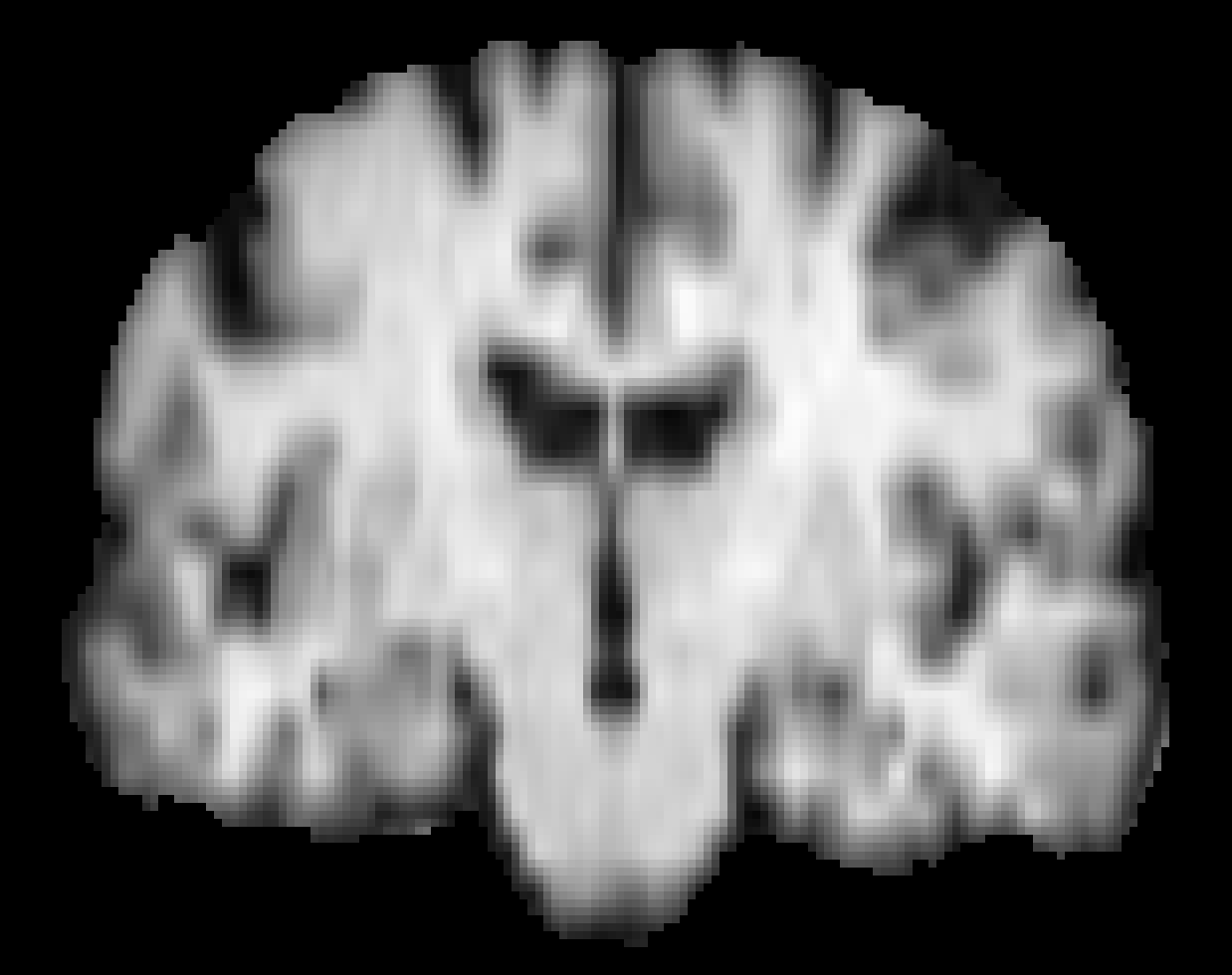}};
            \node[] at (2, 2)[anchor=south west]  {\includegraphics[width=2cm]{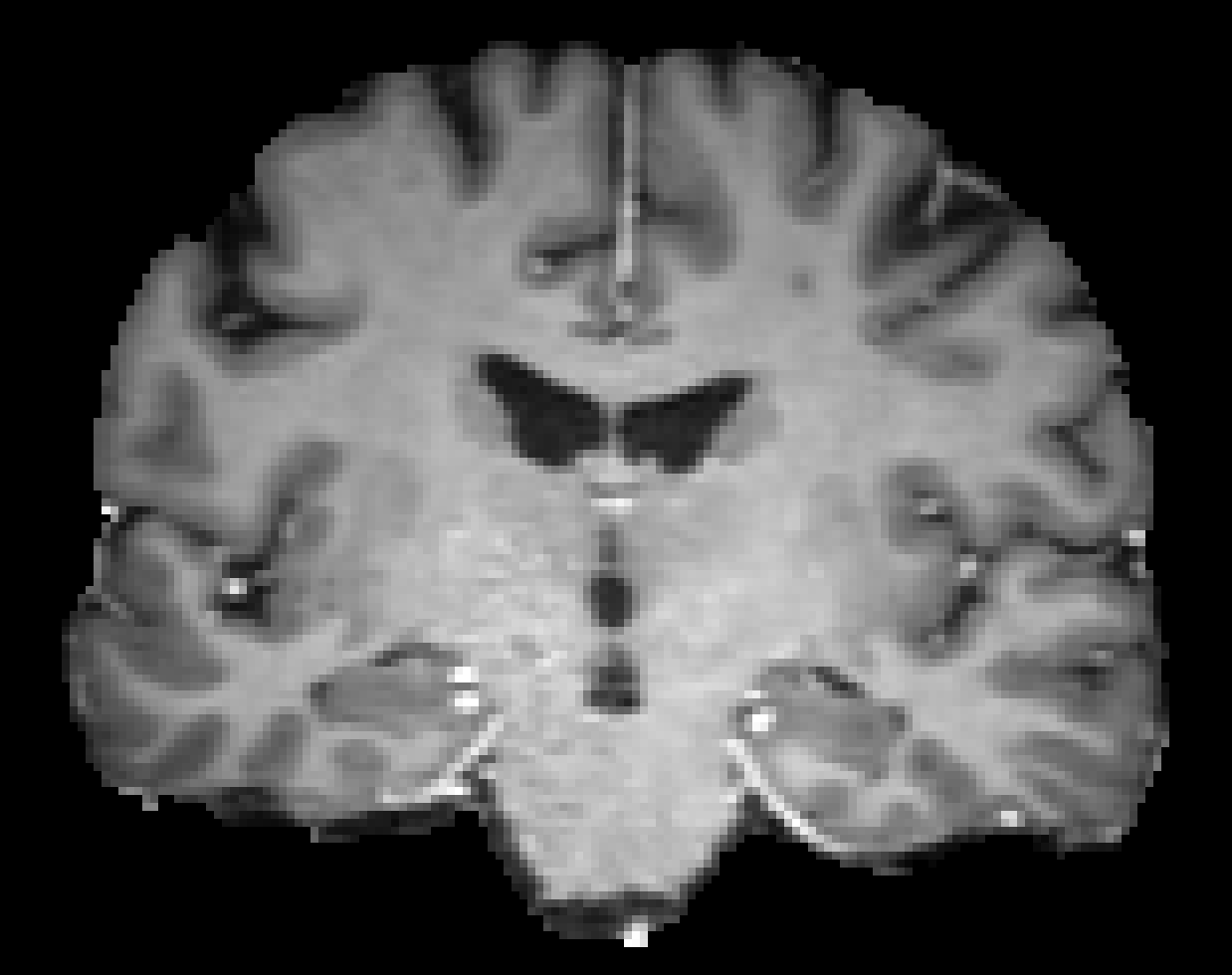}};
            \node[] at (4, 2)[anchor=south west]  {\includegraphics[width=2cm]{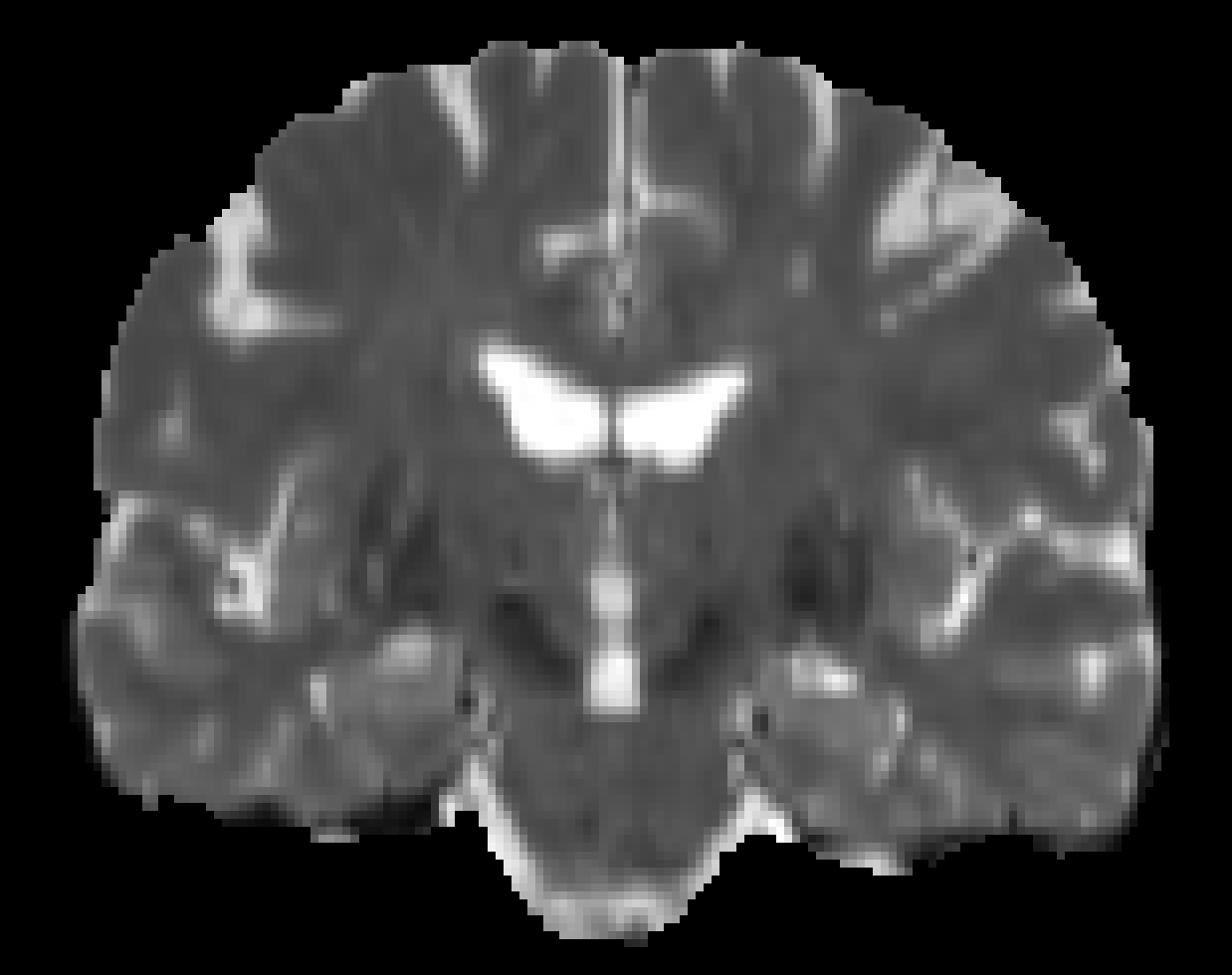}};
            \node[] at (6, 2)[anchor=south west]  {\includegraphics[width=2cm]{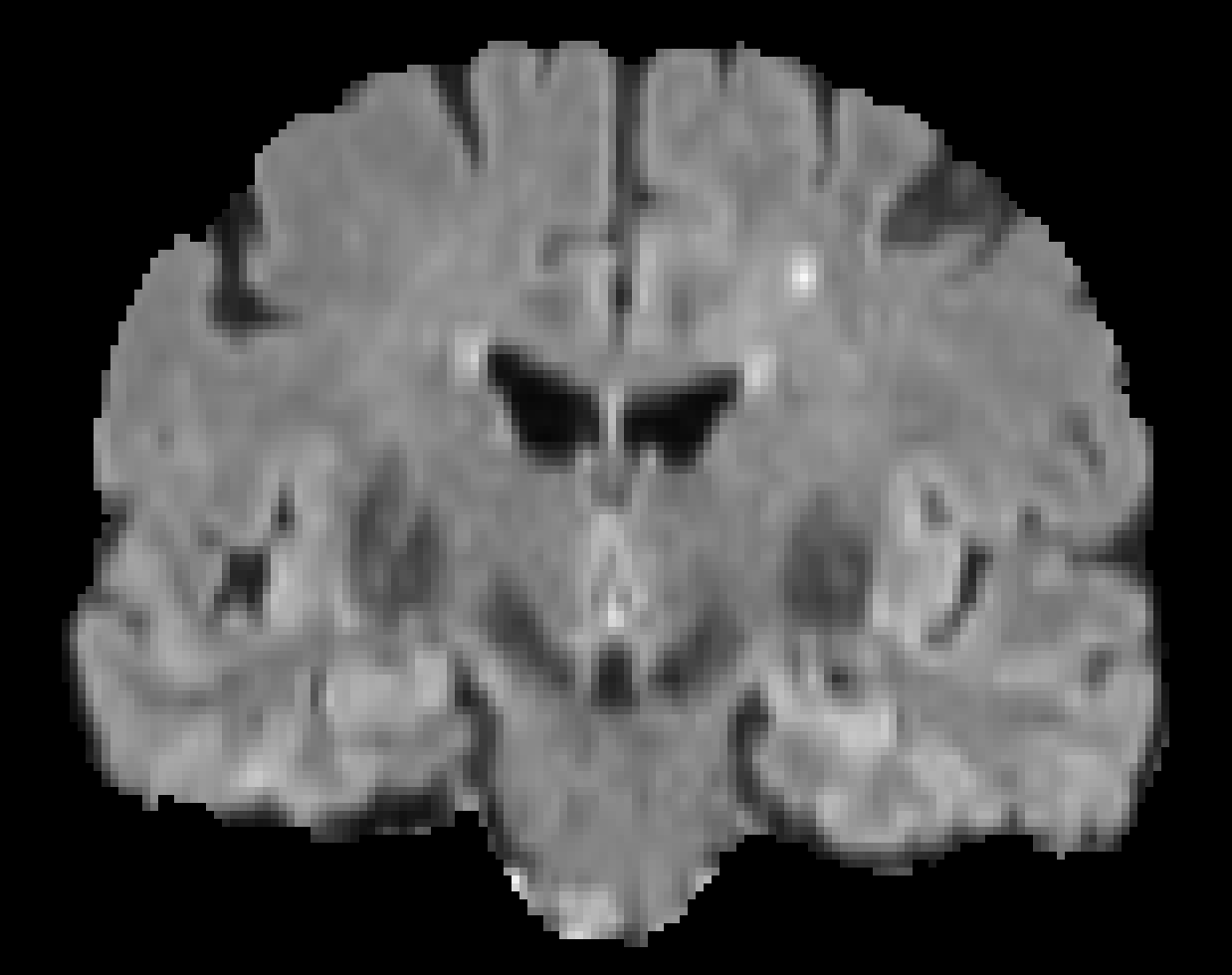}};
            \node[] at (0, 0)[anchor=south west]  {\includegraphics[width=2cm]{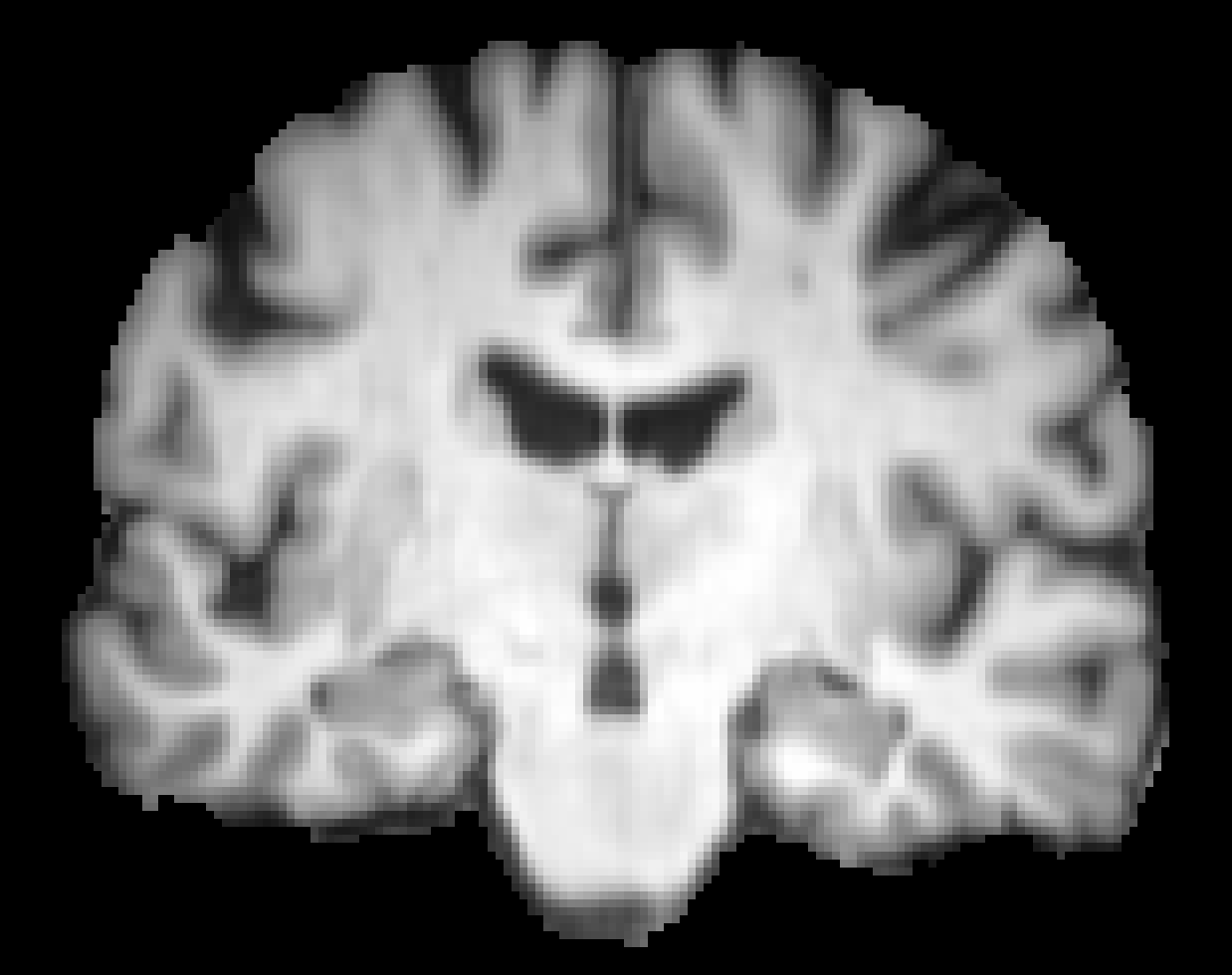}};
            \node[] at (2, 0)[anchor=south west]  {\includegraphics[width=2cm]{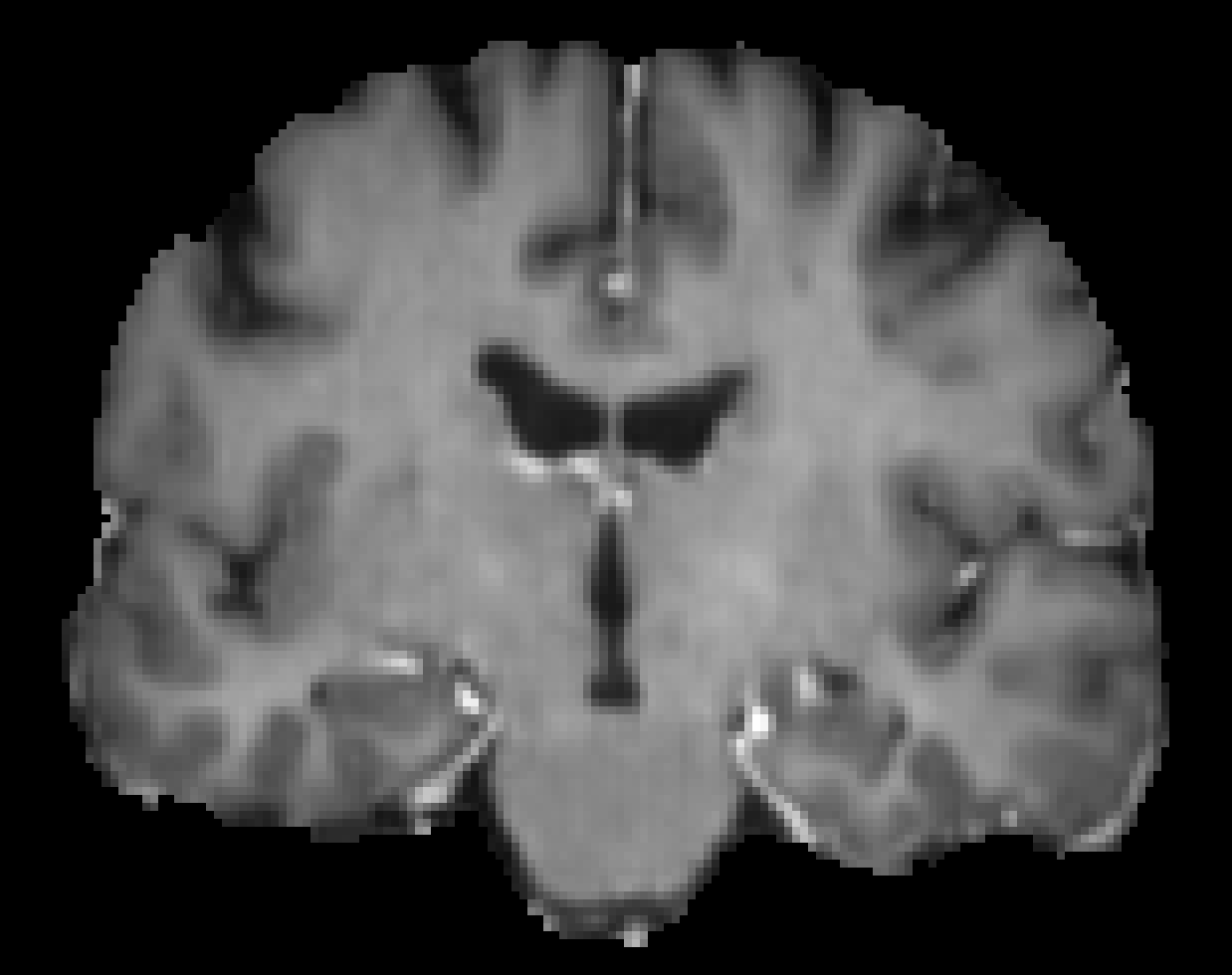}};
            \node[] at (4, 0)[anchor=south west]  {\includegraphics[width=2cm]{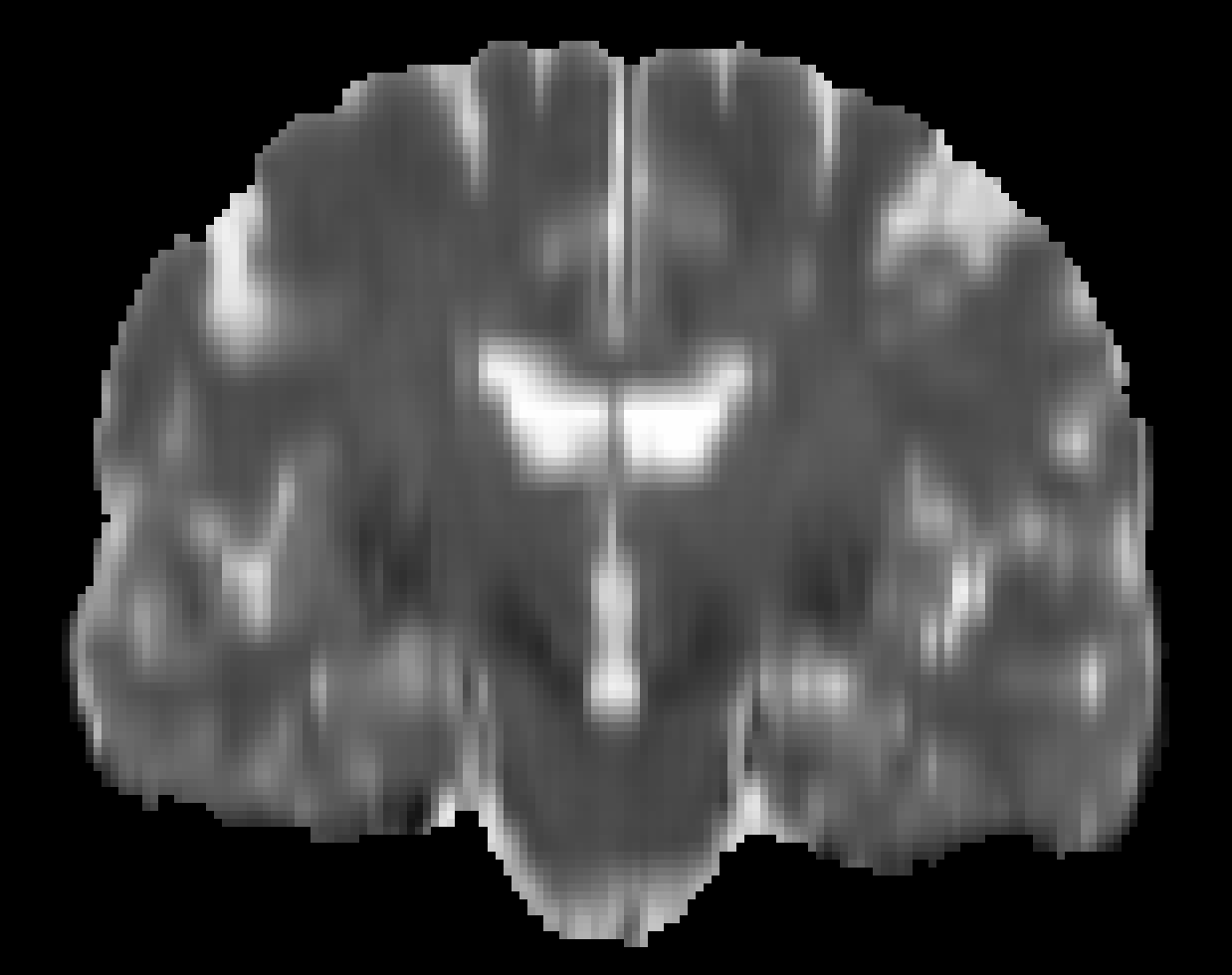}};
            \node[] at (6, 0)[anchor=south west]  {\includegraphics[width=2cm]{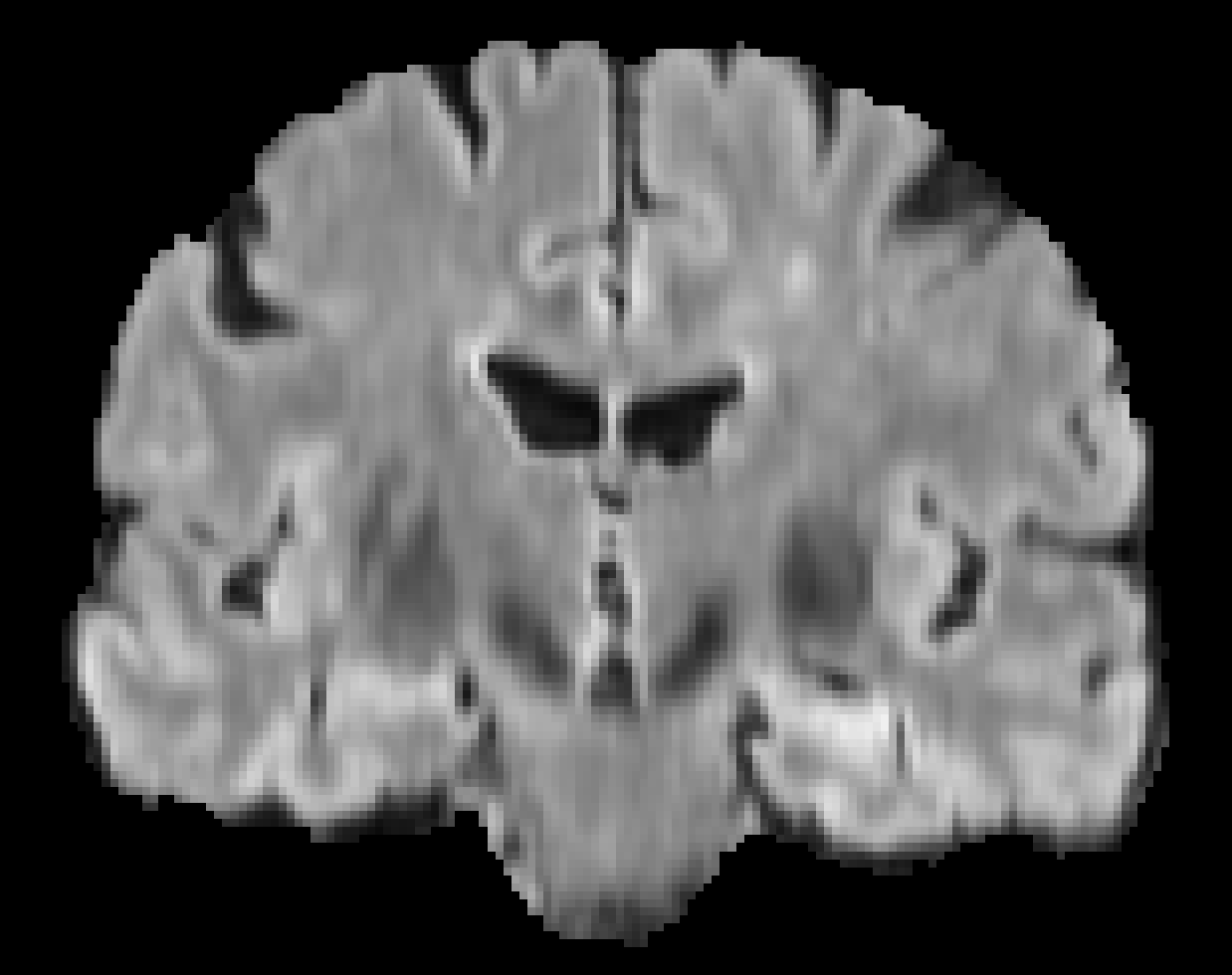}};
            \node[] at (0, 6)[anchor=south west]  {\includegraphics[width=2cm]{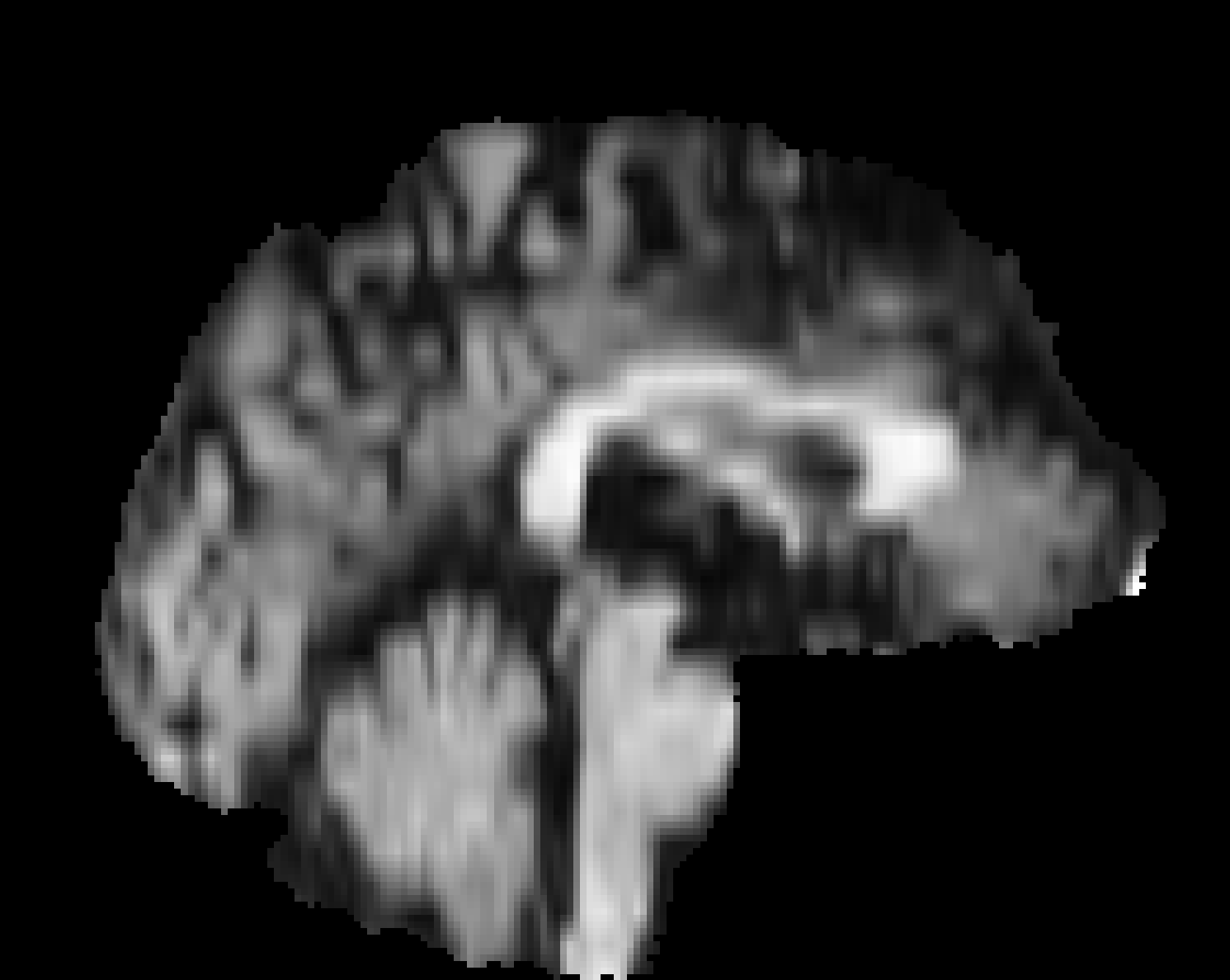}};
            \node[] at (2, 6)[anchor=south west]  {\includegraphics[width=2cm]{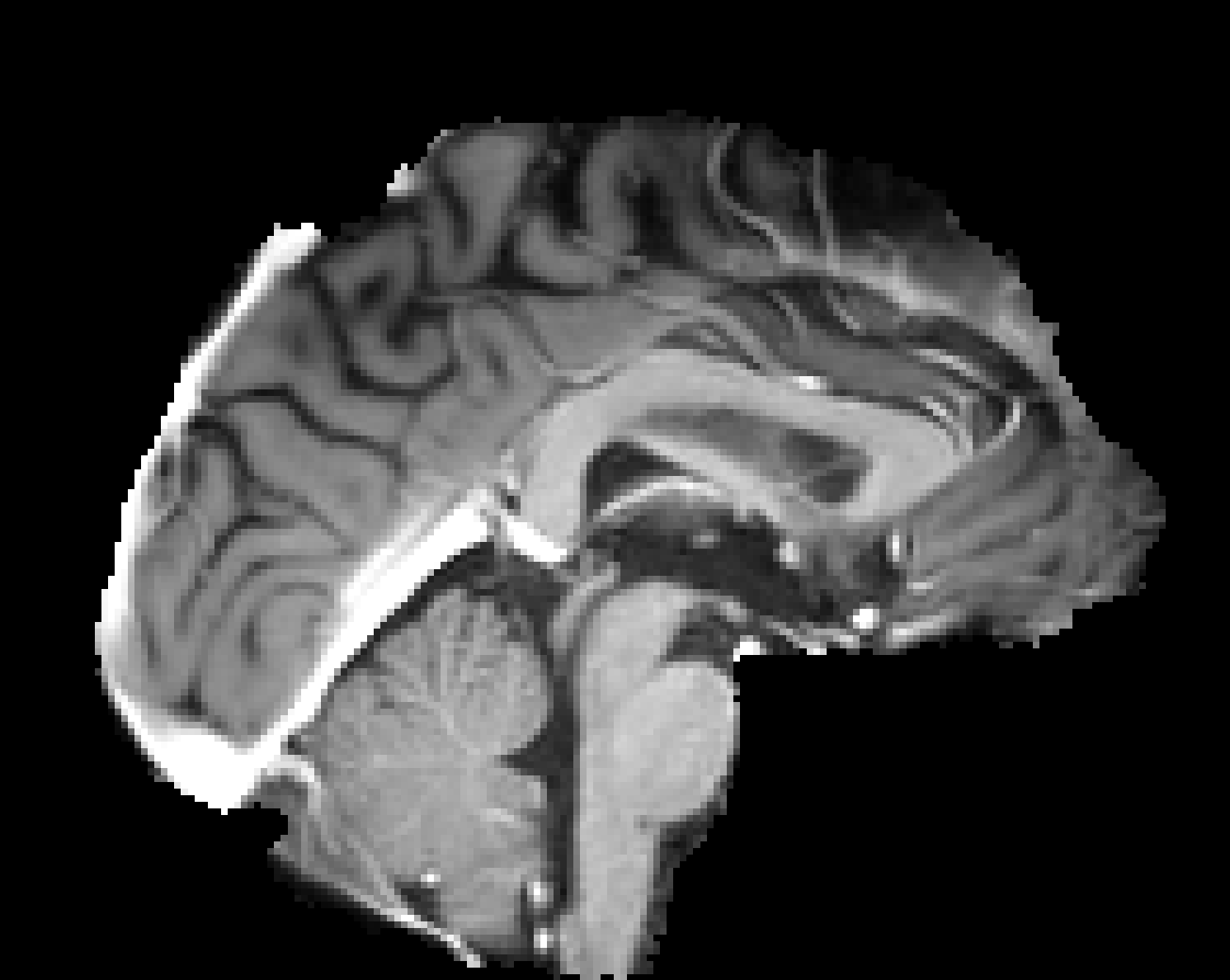}};
            \node[] at (4, 6)[anchor=south west]  {\includegraphics[width=2cm]{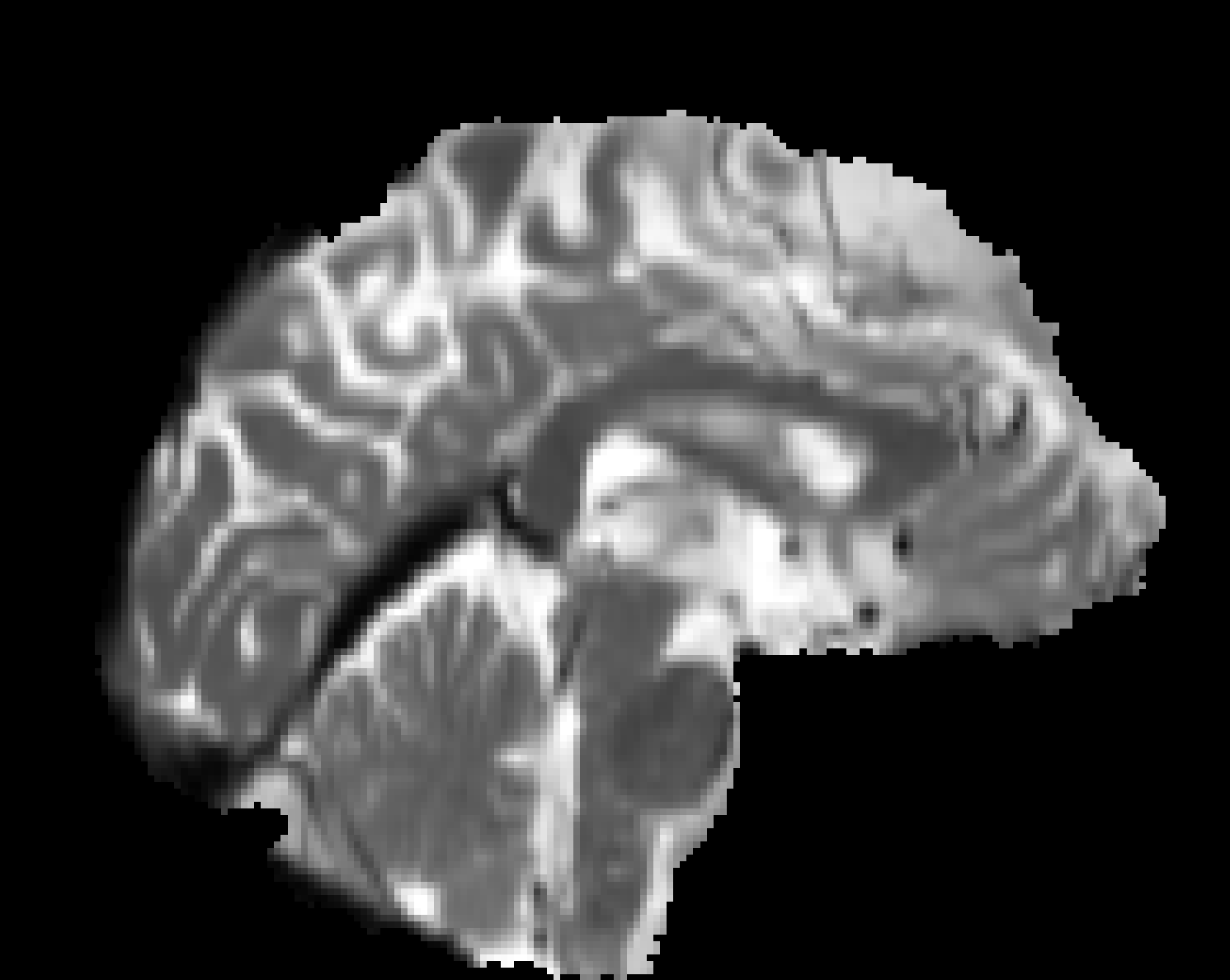}};
            \node[] at (6, 6)[anchor=south west]  {\includegraphics[width=2cm]{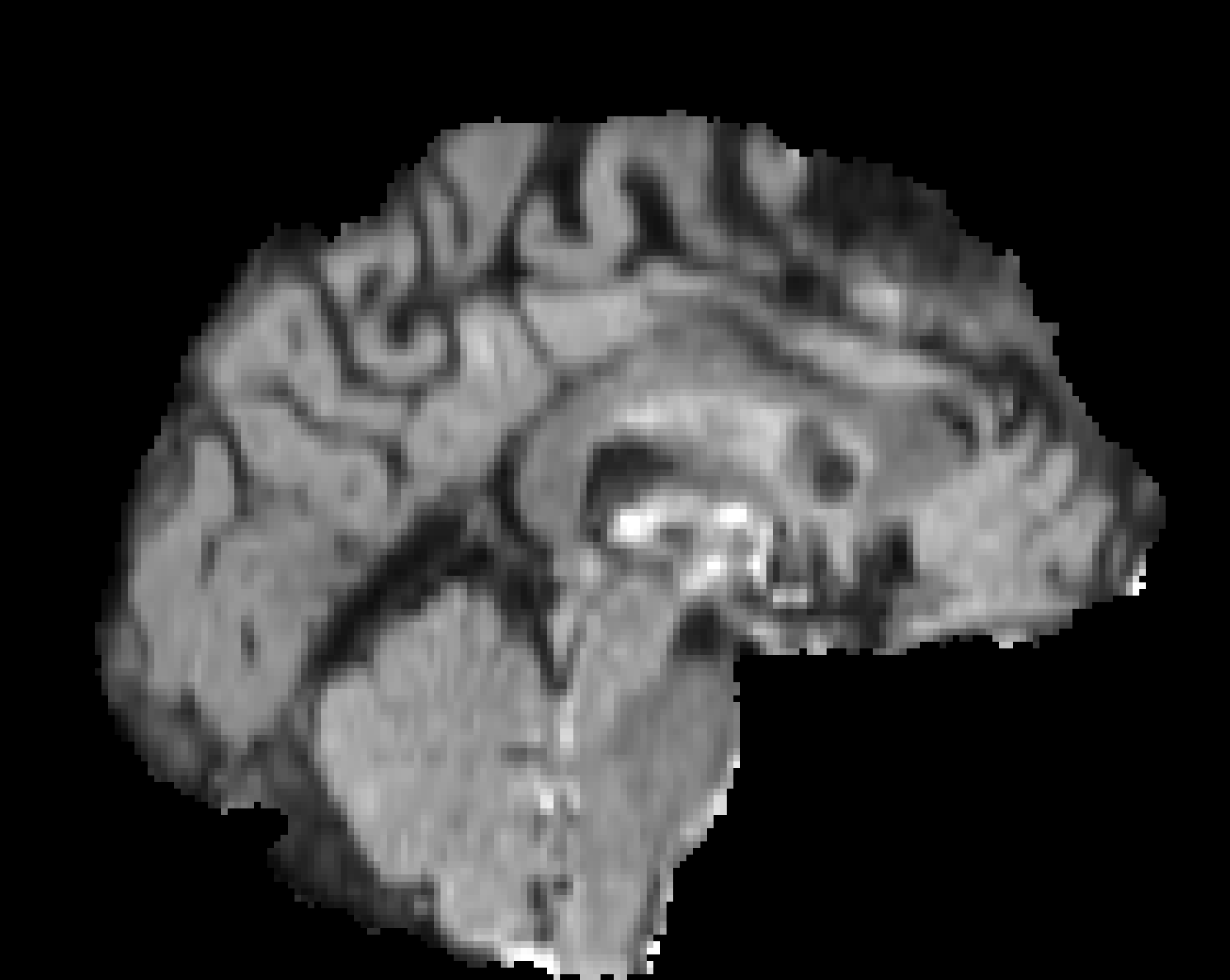}};
            \node[] at (0, 4)[anchor=south west]  {\includegraphics[width=2cm]{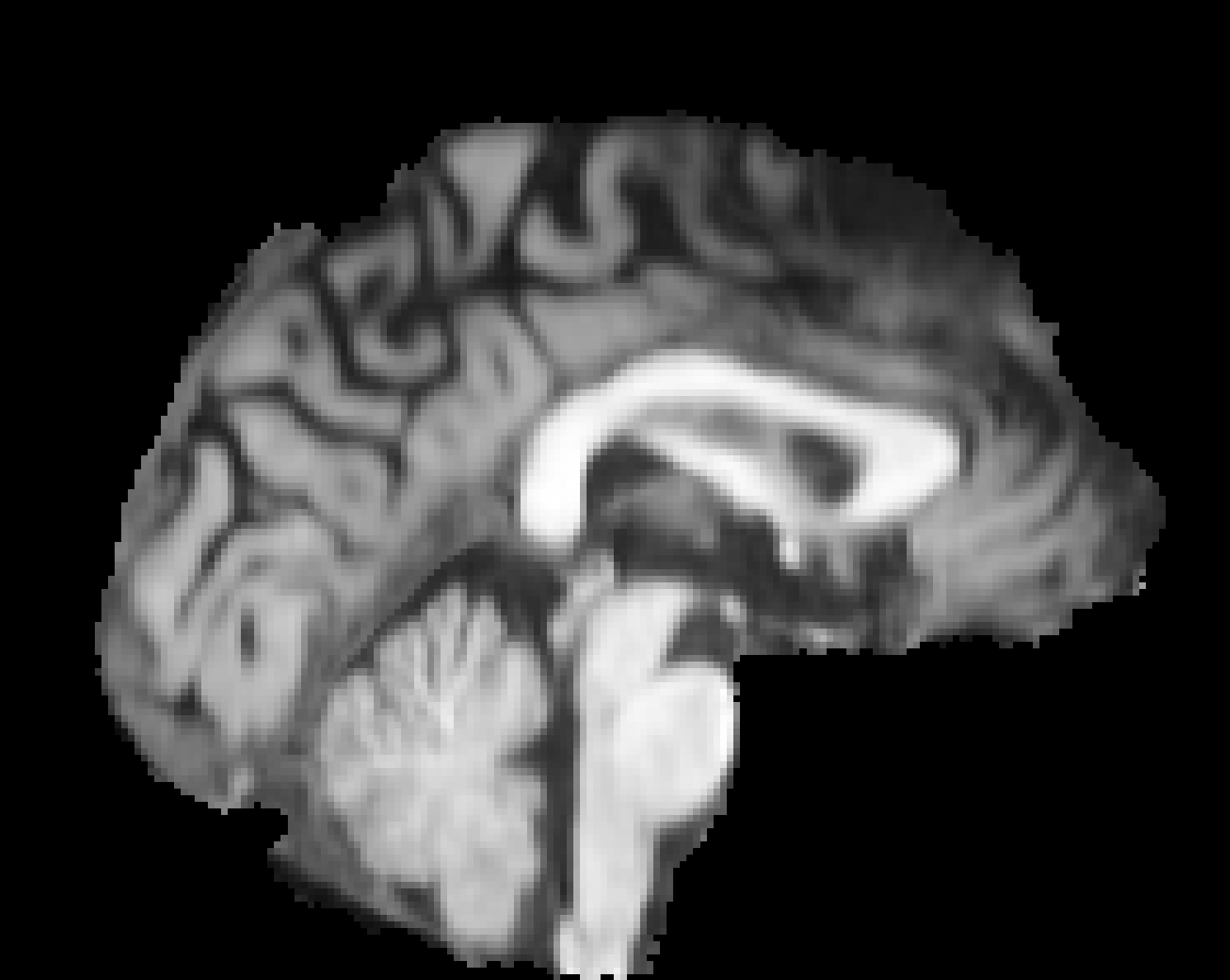}};
            \node[] at (2, 4)[anchor=south west]  {\includegraphics[width=2cm]{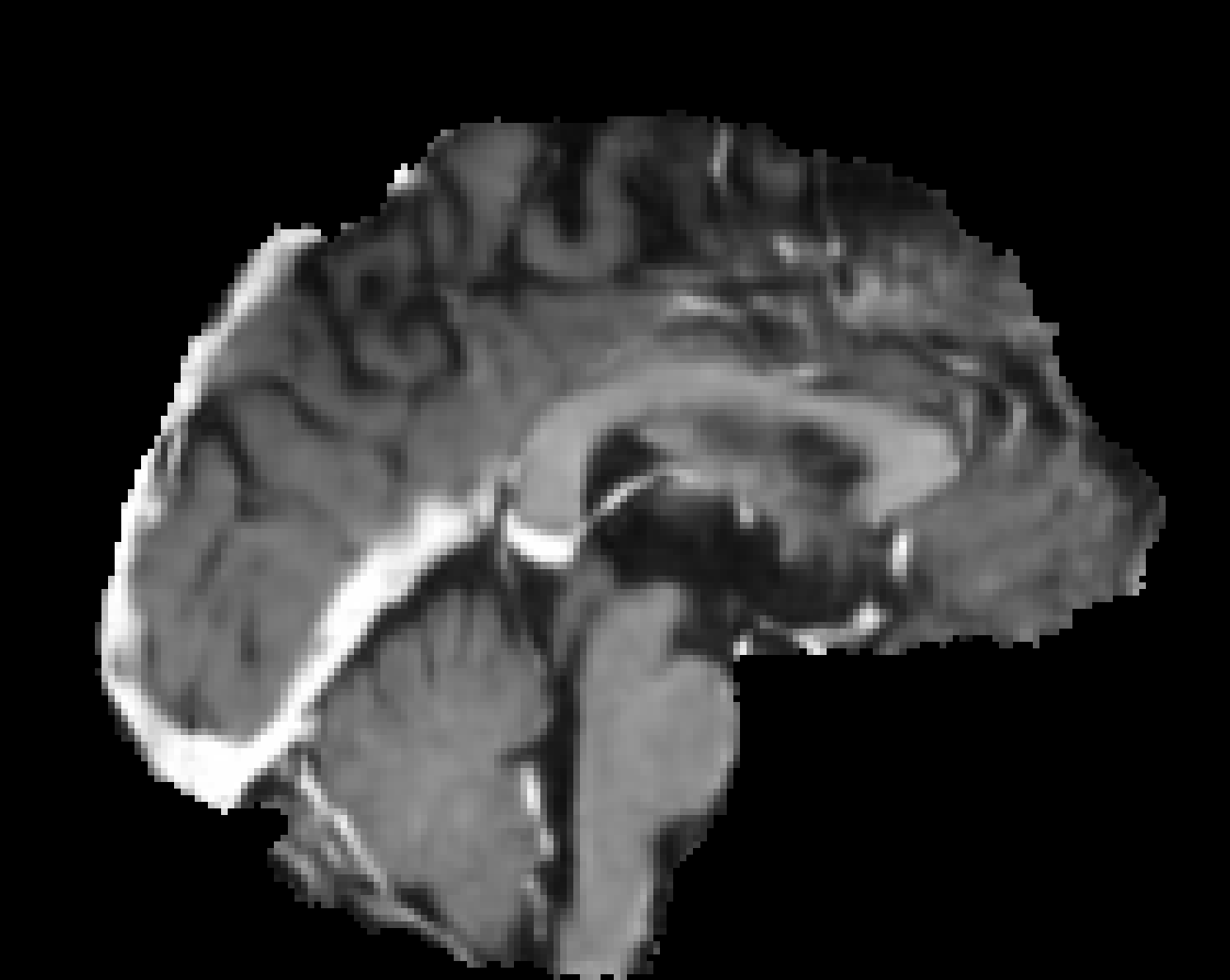}};
            \node[] at (4, 4)[anchor=south west]  {\includegraphics[width=2cm]{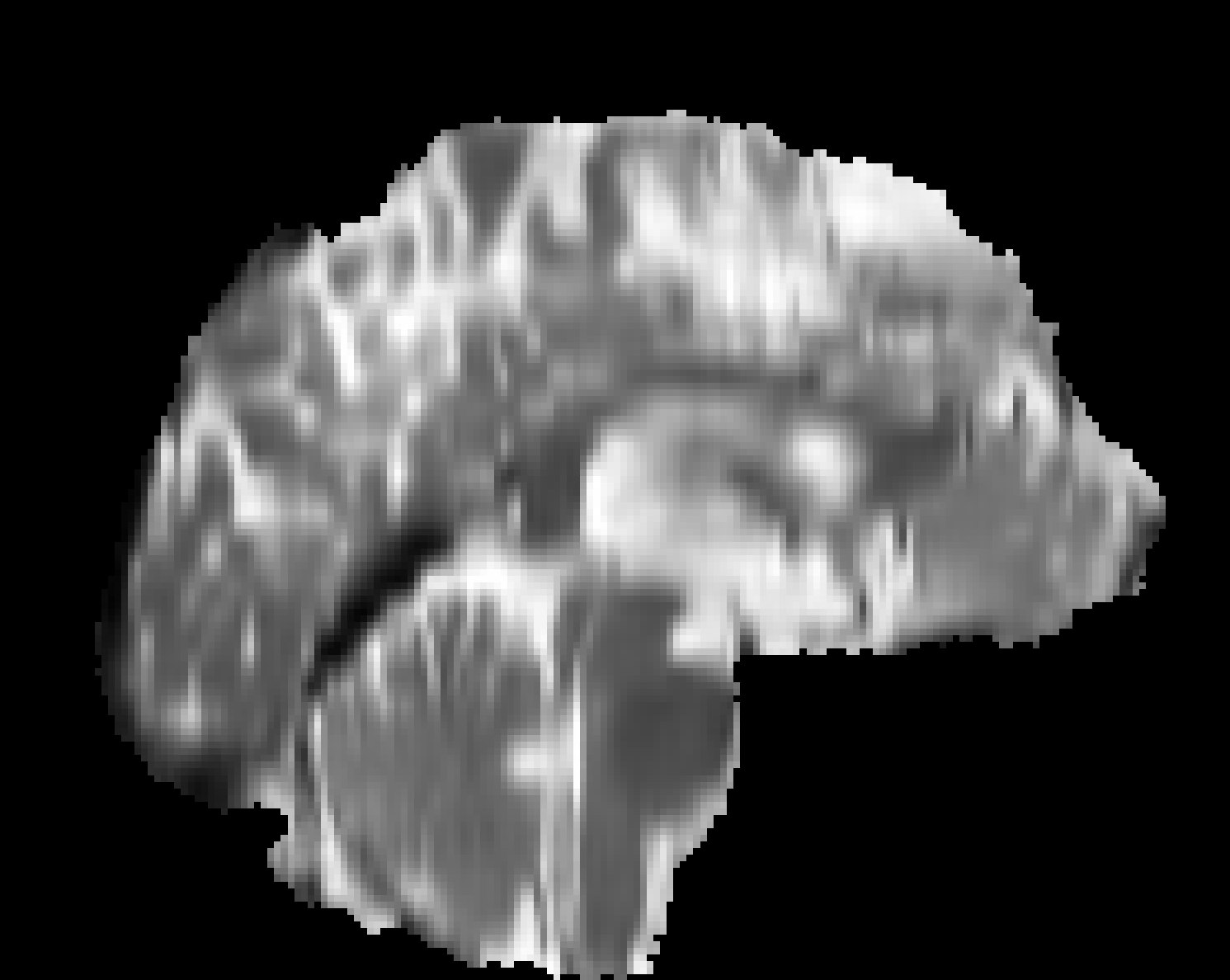}};
            \node[] at (6, 4)[anchor=south west]  {\includegraphics[width=2cm]{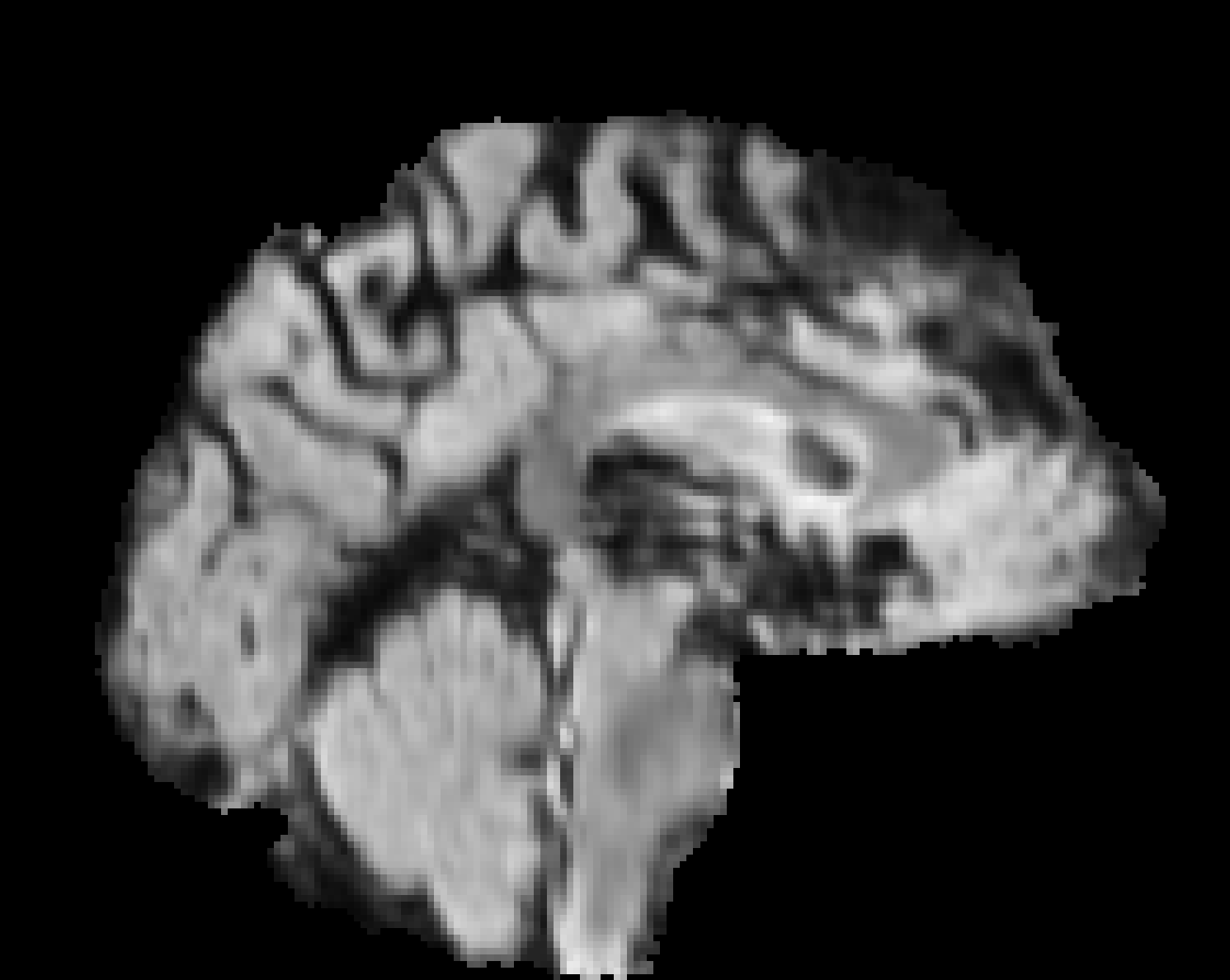}};
            \node[] at (0, 10)[anchor=south west]  {\includegraphics[width=2cm]{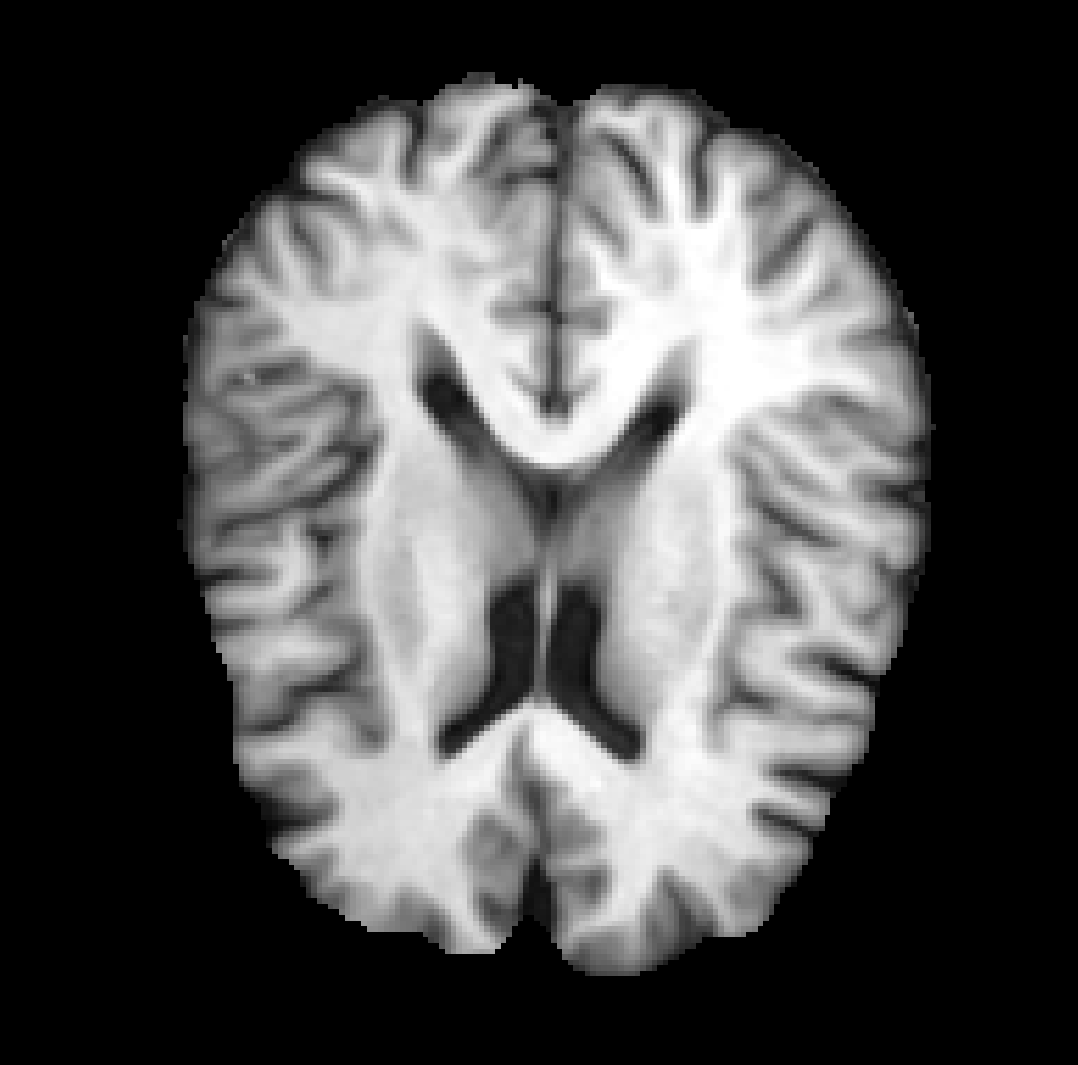}};
            \node[] at (2, 10)[anchor=south west]  {\includegraphics[width=2cm]{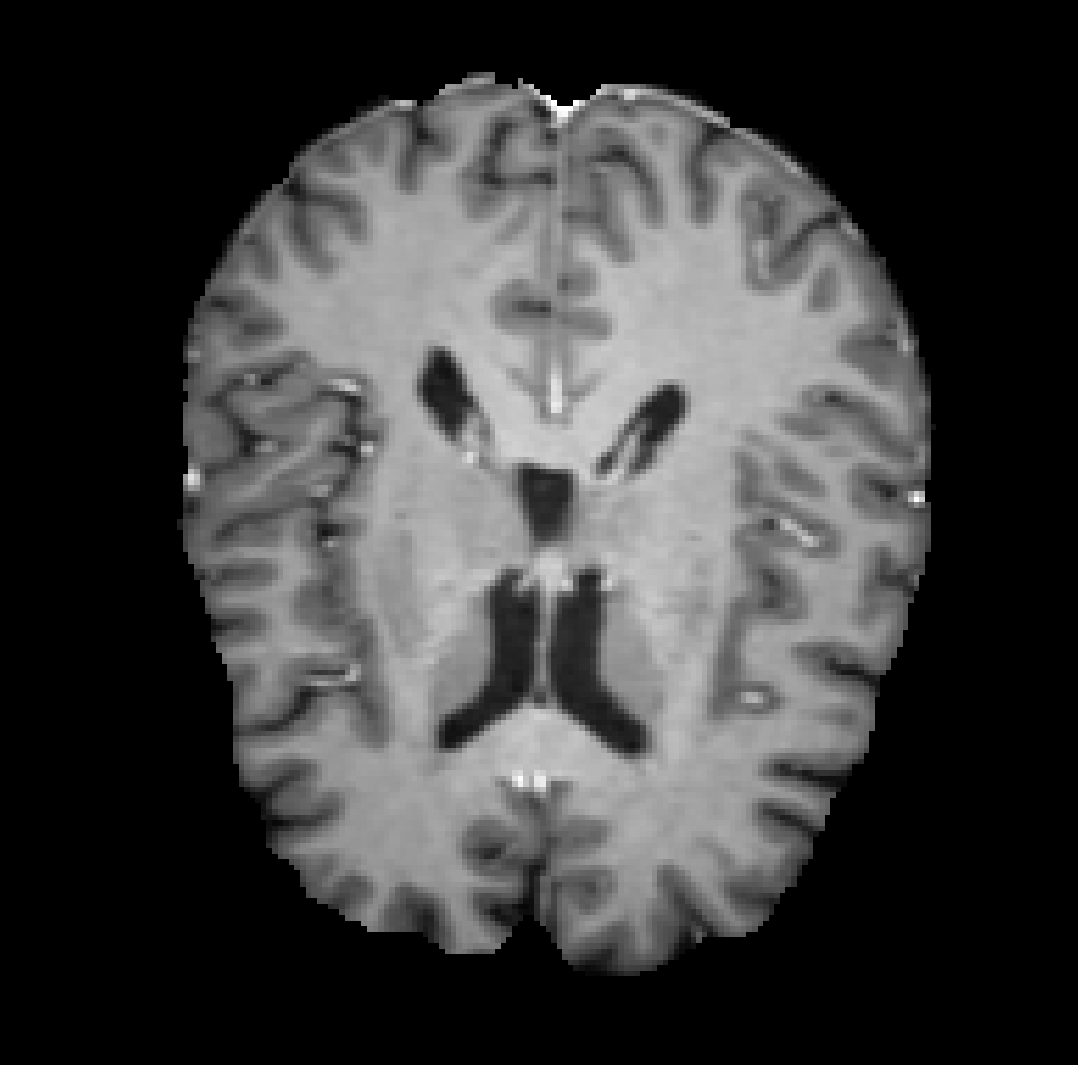}};
            \node[] at (4, 10)[anchor=south west]  {\includegraphics[width=2cm]{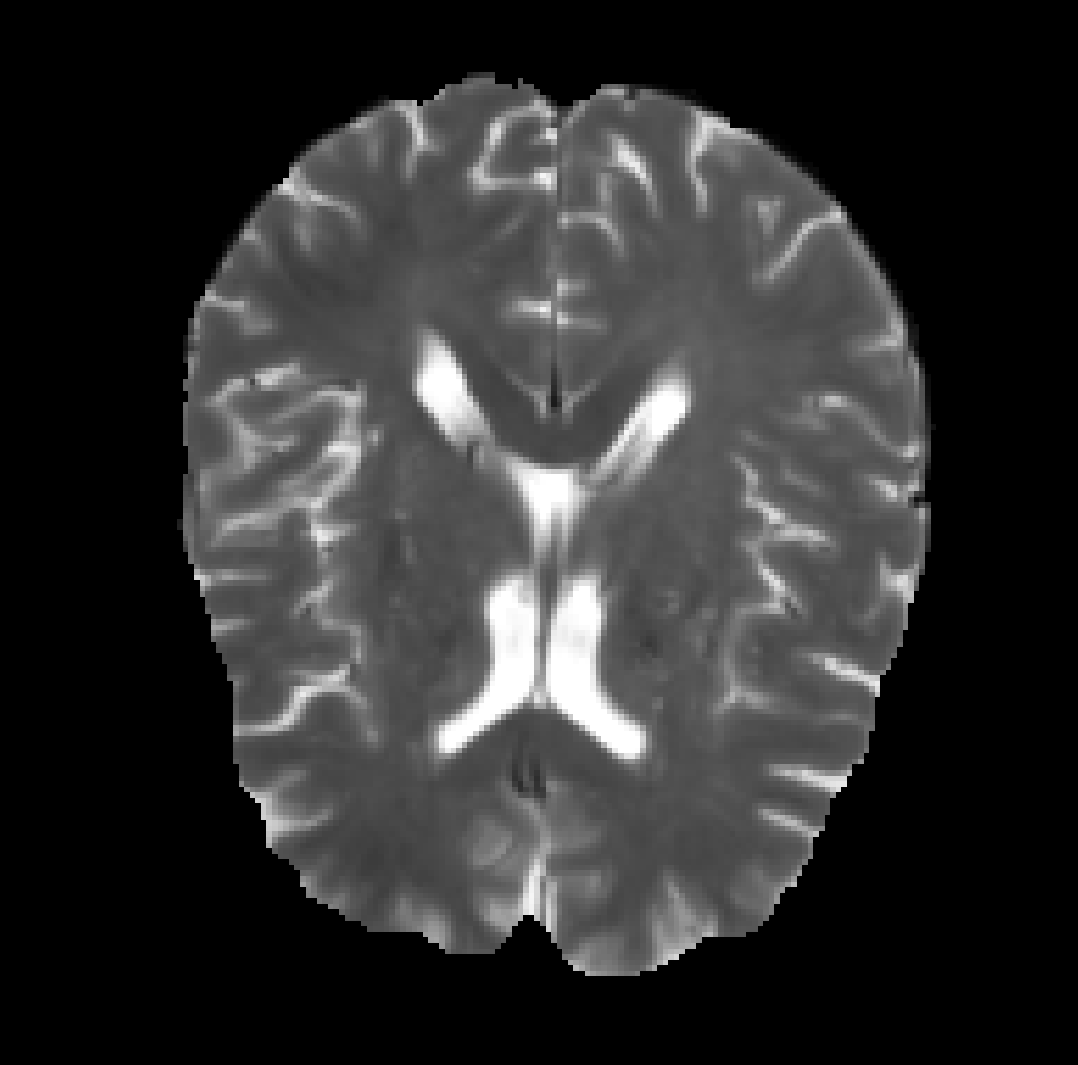}};
            \node[] at (6, 10)[anchor=south west]  {\includegraphics[width=2cm]{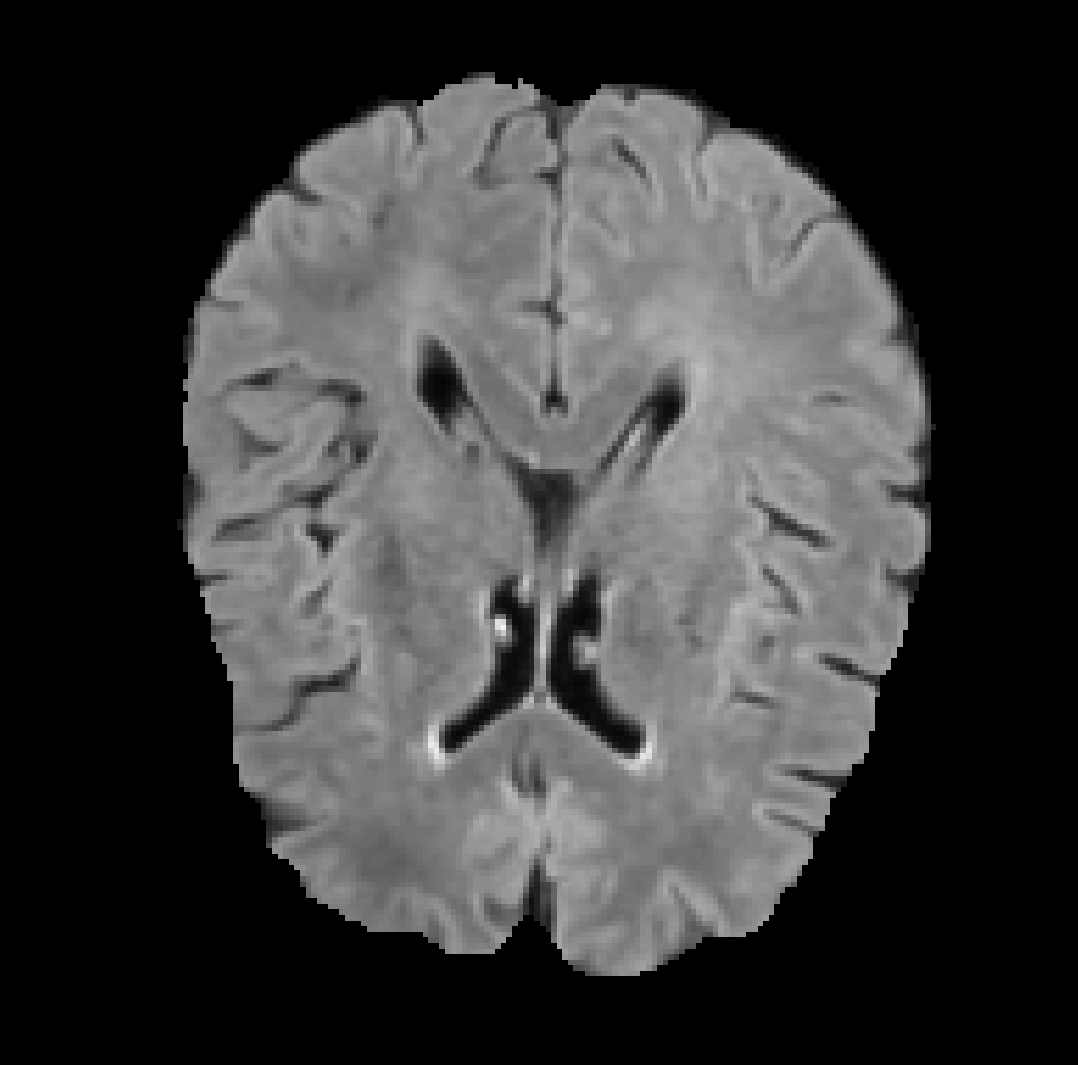}};
            \node[] at (0, 8)[anchor=south west]  {\includegraphics[width=2cm]{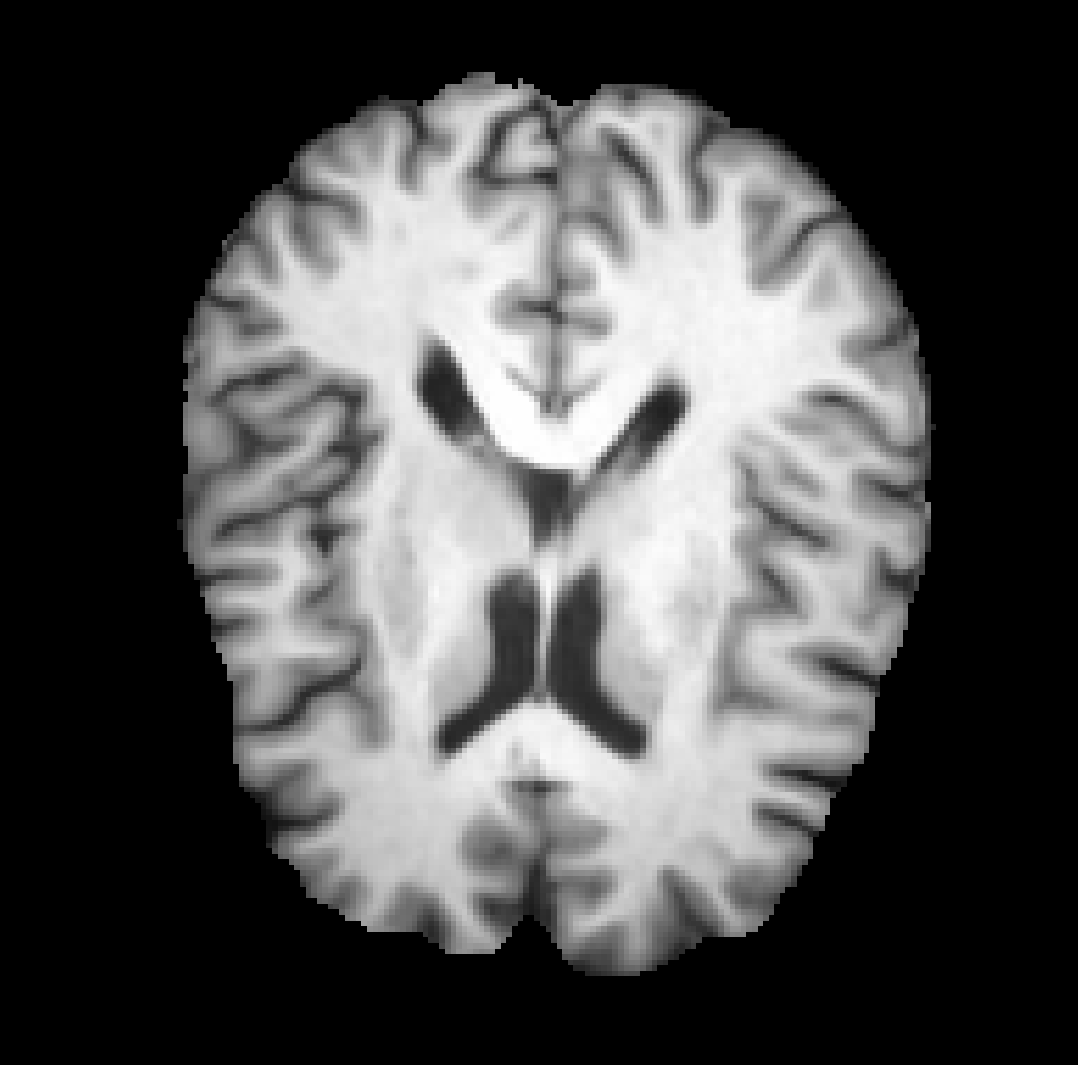}};
            \node[] at (2, 8)[anchor=south west]  {\includegraphics[width=2cm]{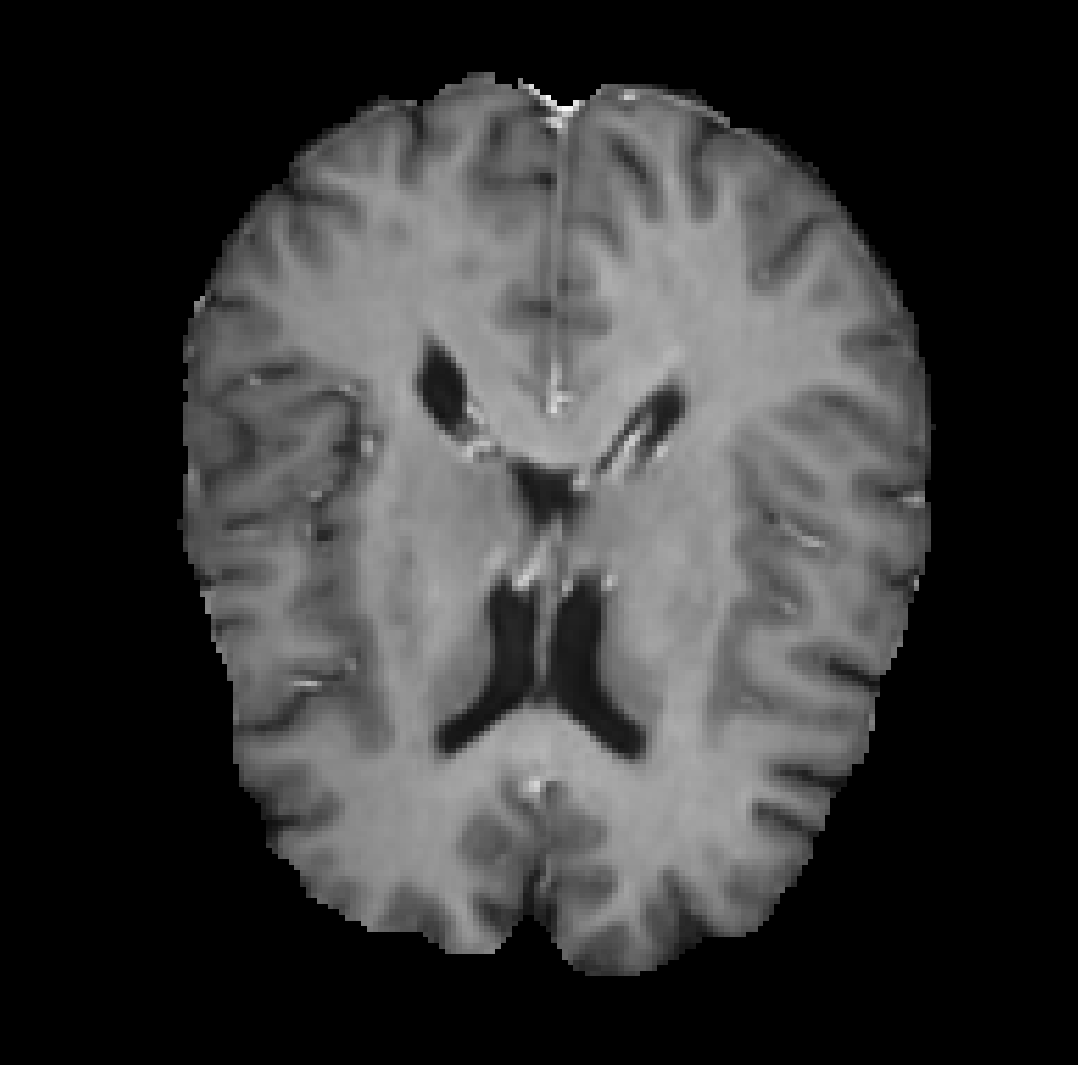}};
            \node[] at (4, 8)[anchor=south west]  {\includegraphics[width=2cm]{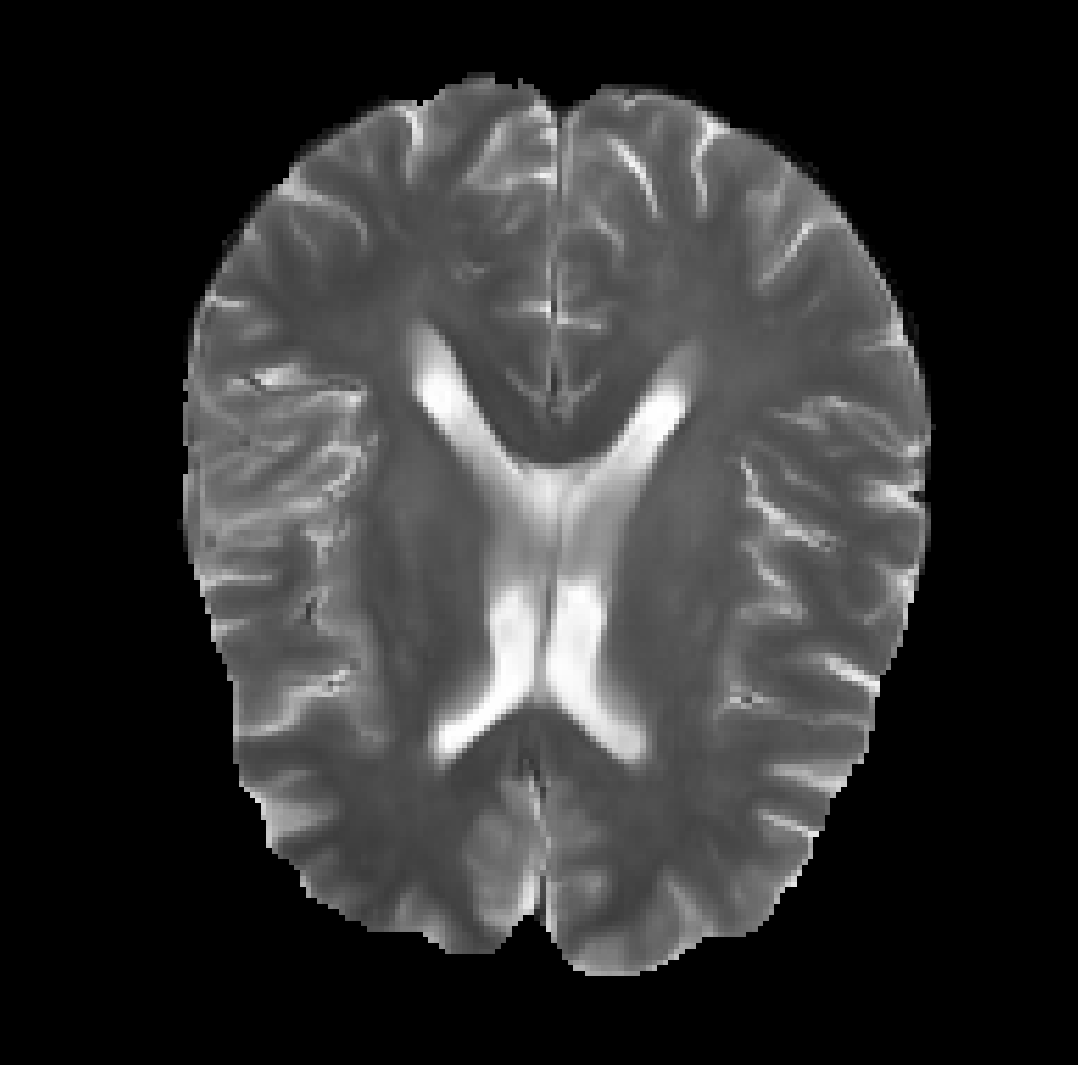}};
            \node[] at (6, 8)[anchor=south west]  {\includegraphics[width=2cm]{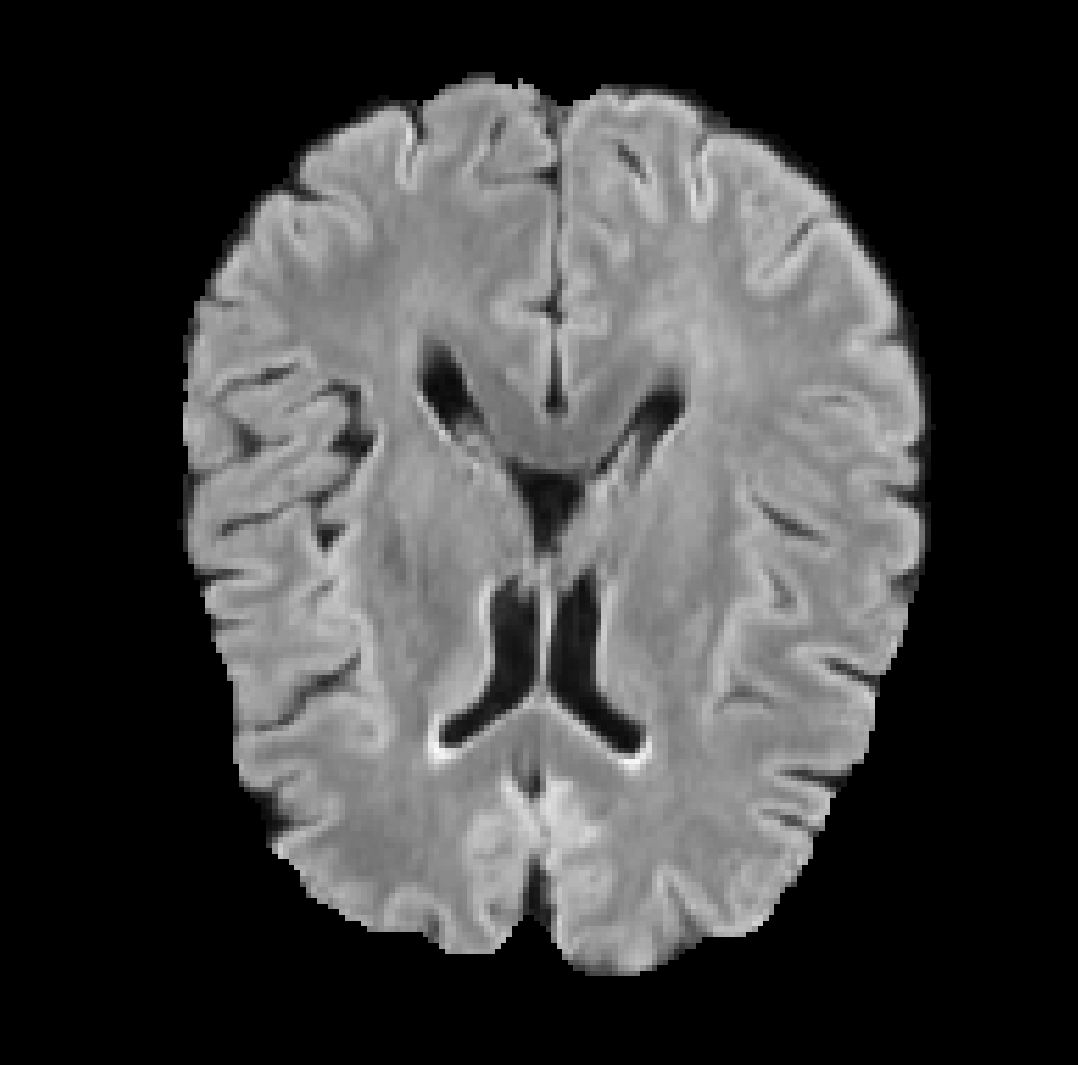}};
            \node[rotate=90] at (-0.2, 1)            {\scriptsize Synthetic};
            \node[rotate=90] at (-0.2, 3)            {\scriptsize Real};
            \node[rotate=90] at (-0.2, 5)            {\scriptsize Synthetic};
            \node[rotate=90] at (-0.2, 7)            {\scriptsize Real};
            \node[rotate=90] at (-0.2, 9)            {\scriptsize Synthetic};
            \node[rotate=90] at (-0.2, 11)            {\scriptsize Real};
            \node[] at (1,12.4) {\scriptsize T1};
            \node[] at (3,12.4) {\scriptsize T1ce};
            \node[] at (5,12.4) {\scriptsize T2};
            \node[] at (7,12.4) {\scriptsize FLAIR};
    	\end{tikzpicture}
     }
    \caption{Qualitative results of our proposed method. The synthetic images are generated conditioned on the real images from the three other modalities. We display the middle slice in the axial \emph{(top)}, sagittal \emph{(middle)}, and coronal \emph{(bottom)} plane.}
    \label{fig:results_1}
\end{figure}

\begin{figure}
    \centering
    \resizebox{\textwidth}{!}{
        \begin{tikzpicture}
            \fill (0.1, 0.1) rectangle (8.125, 12.125);
            \node[] at (0, 2)[anchor=south west]  {\includegraphics[width=2cm]{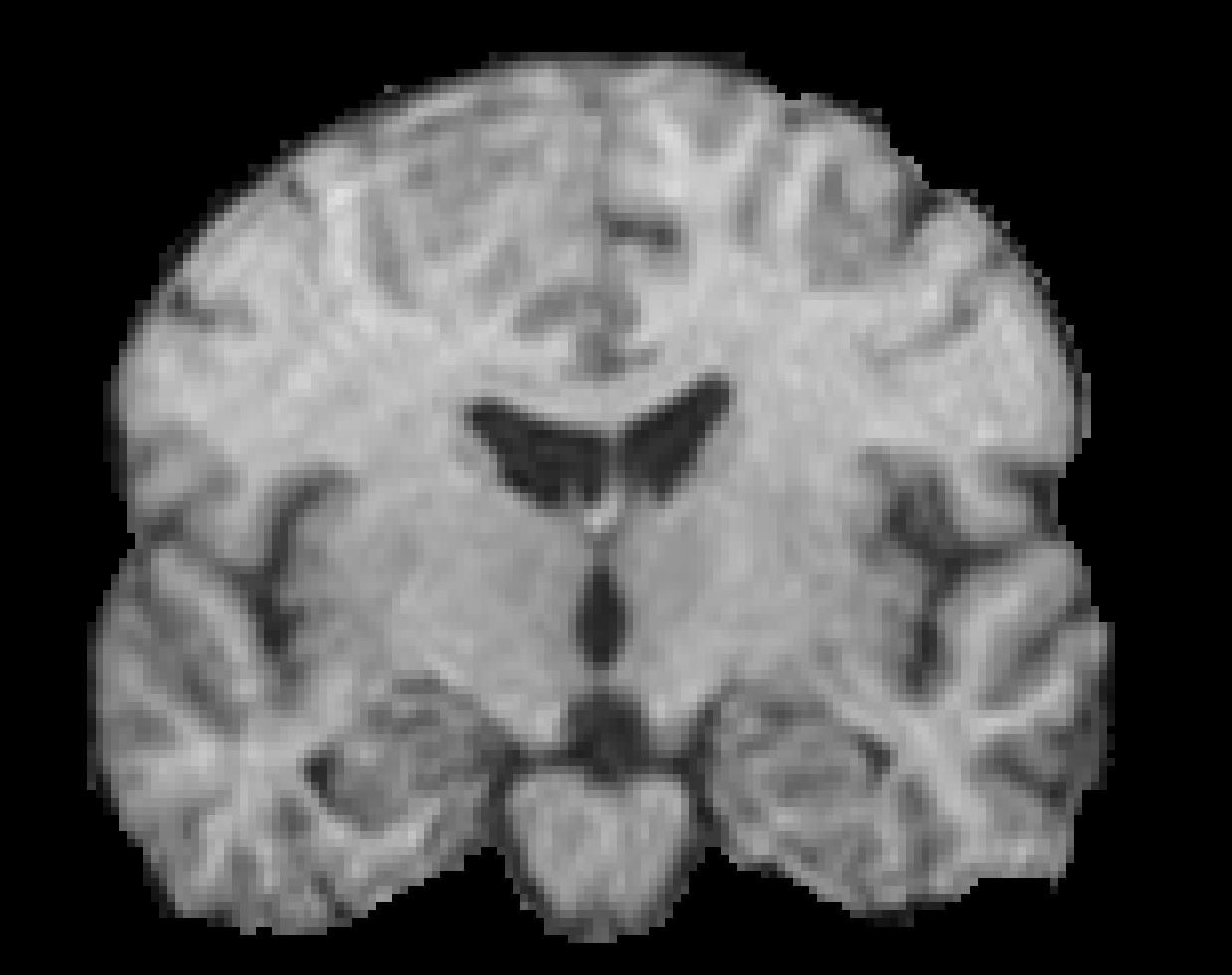}};
            \node[] at (2, 2)[anchor=south west]  {\includegraphics[width=2cm]{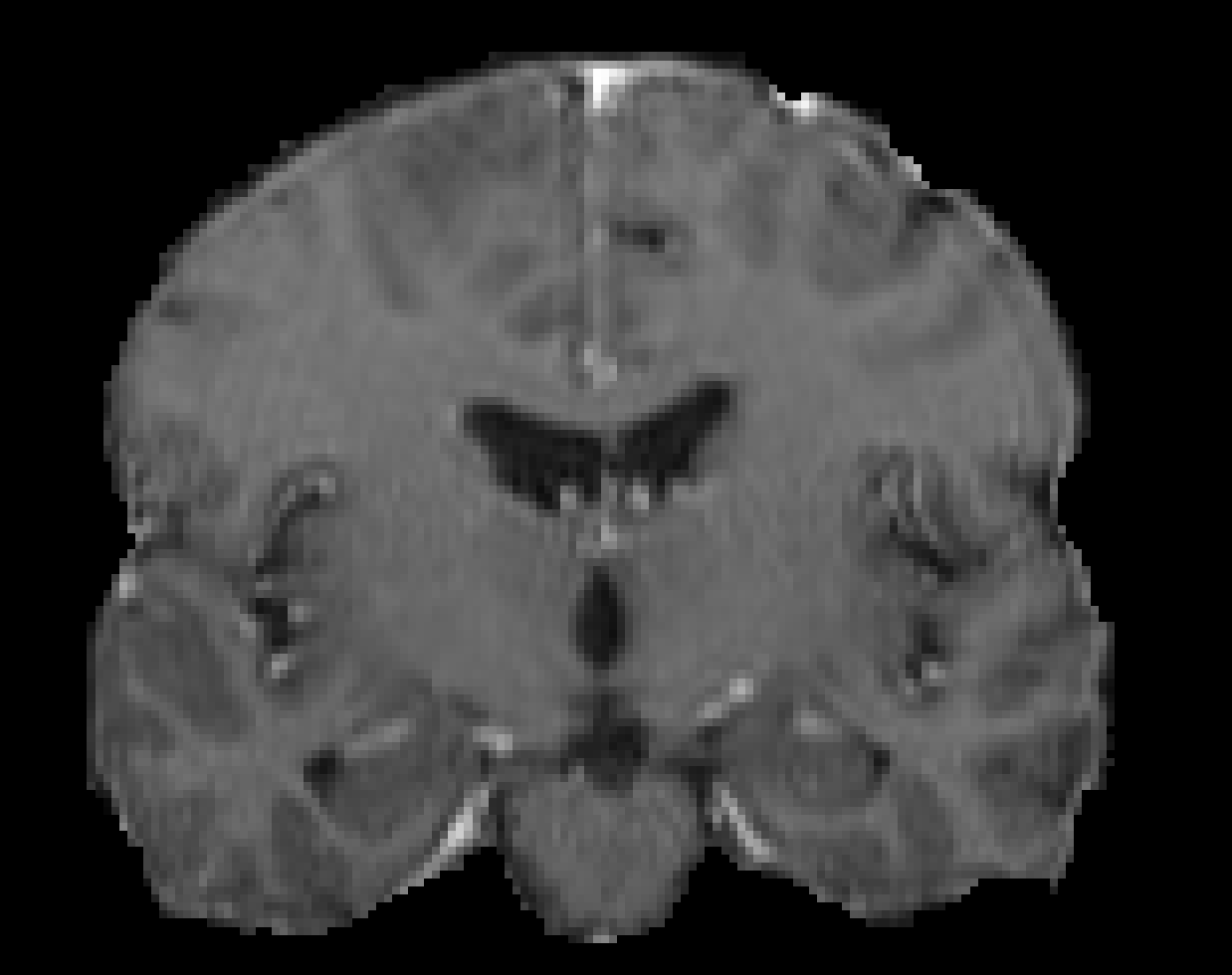}};
            \node[] at (4, 2)[anchor=south west]  {\includegraphics[width=2cm]{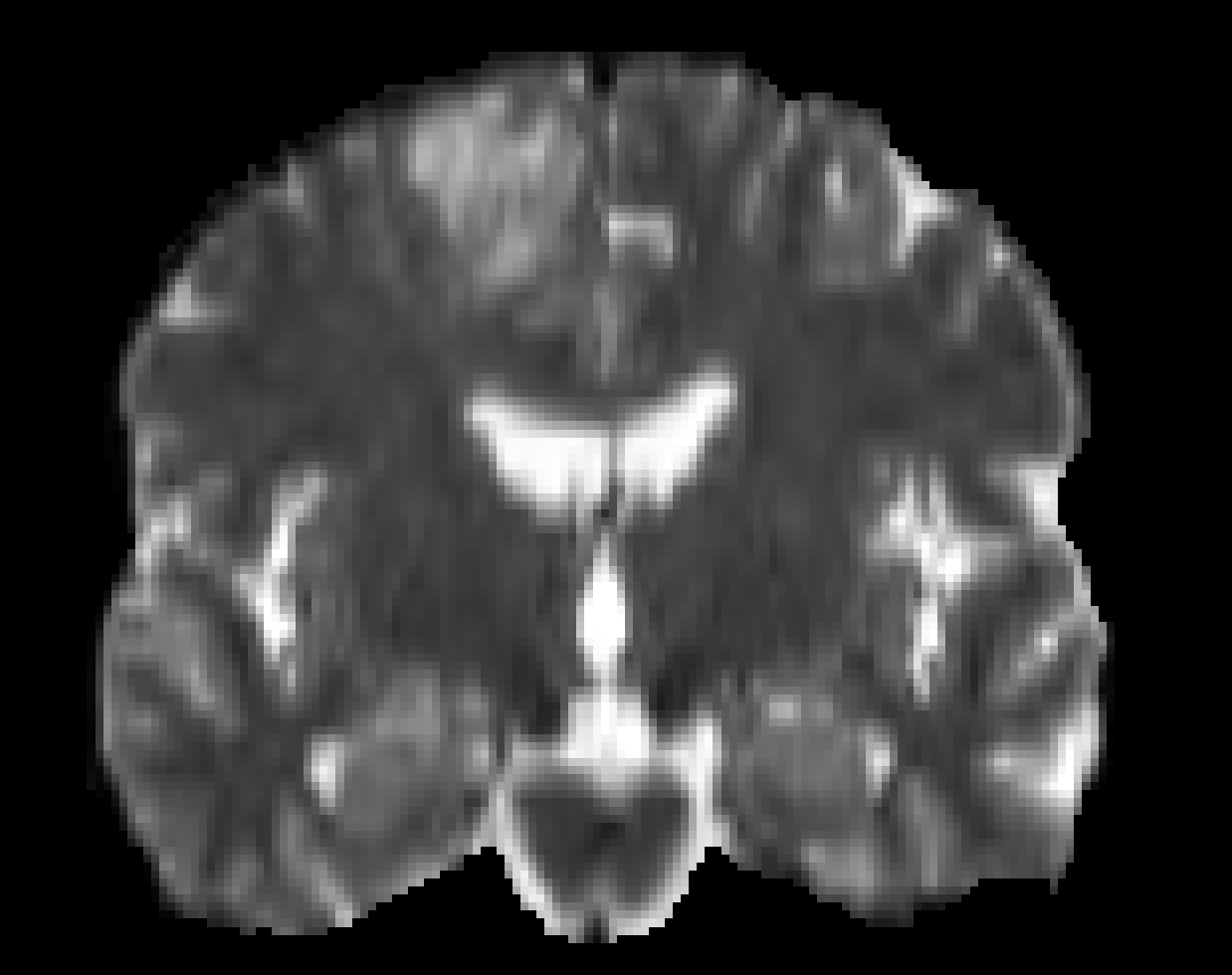}};
            \node[] at (6, 2)[anchor=south west]  {\includegraphics[width=2cm]{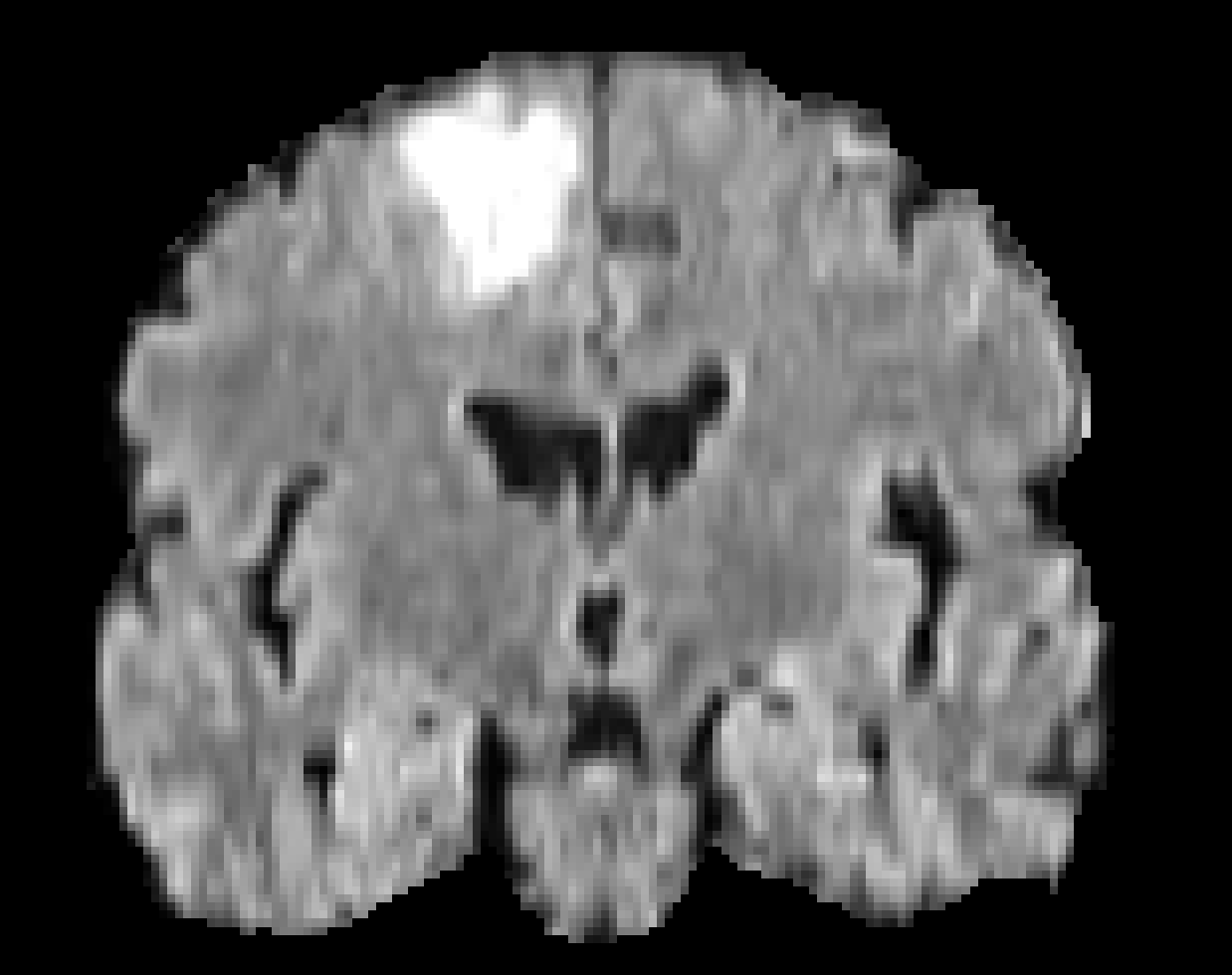}};
            \node[] at (0, 0)[anchor=south west]  {\includegraphics[width=2cm]{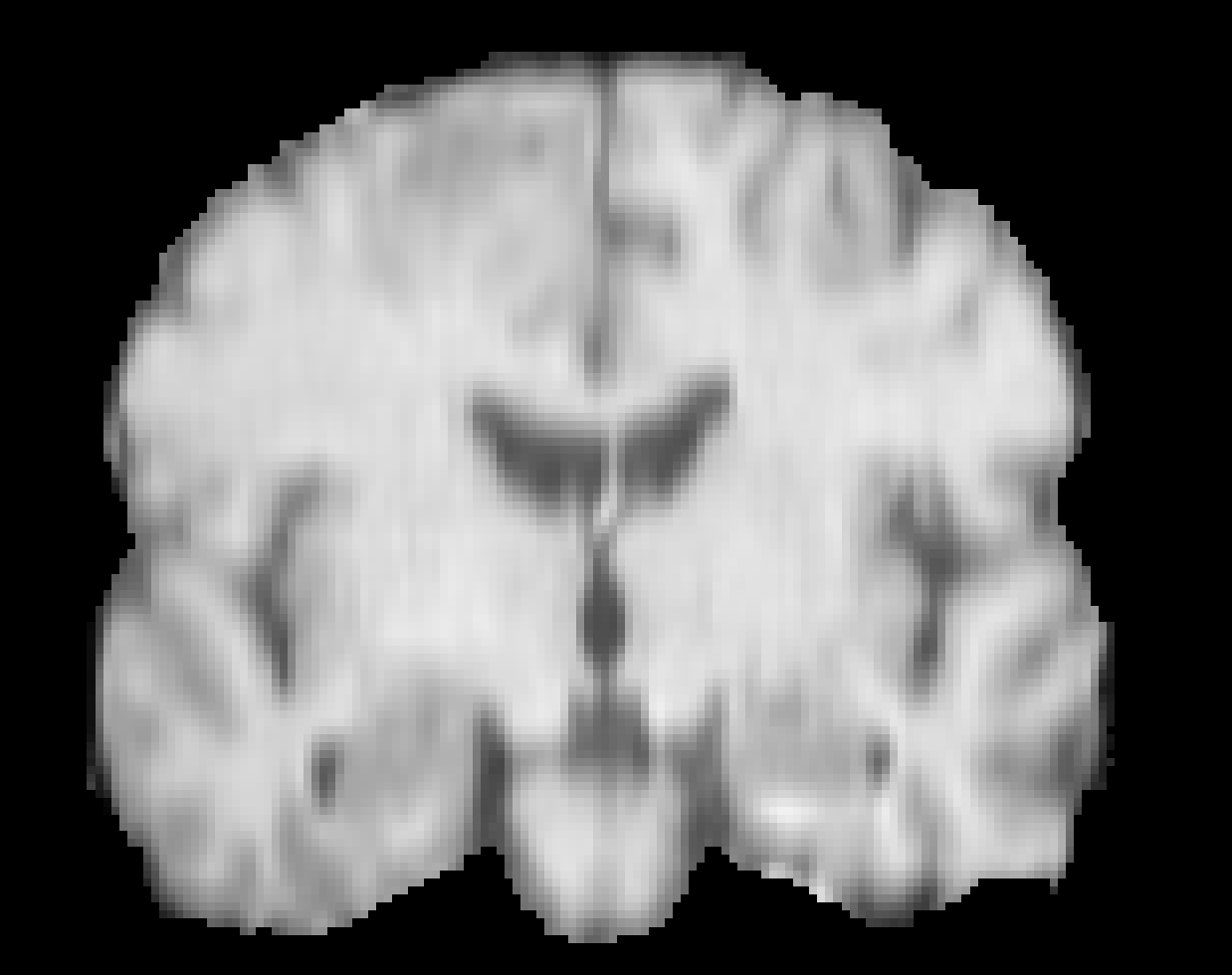}};
            \node[] at (2, 0)[anchor=south west]  {\includegraphics[width=2cm]{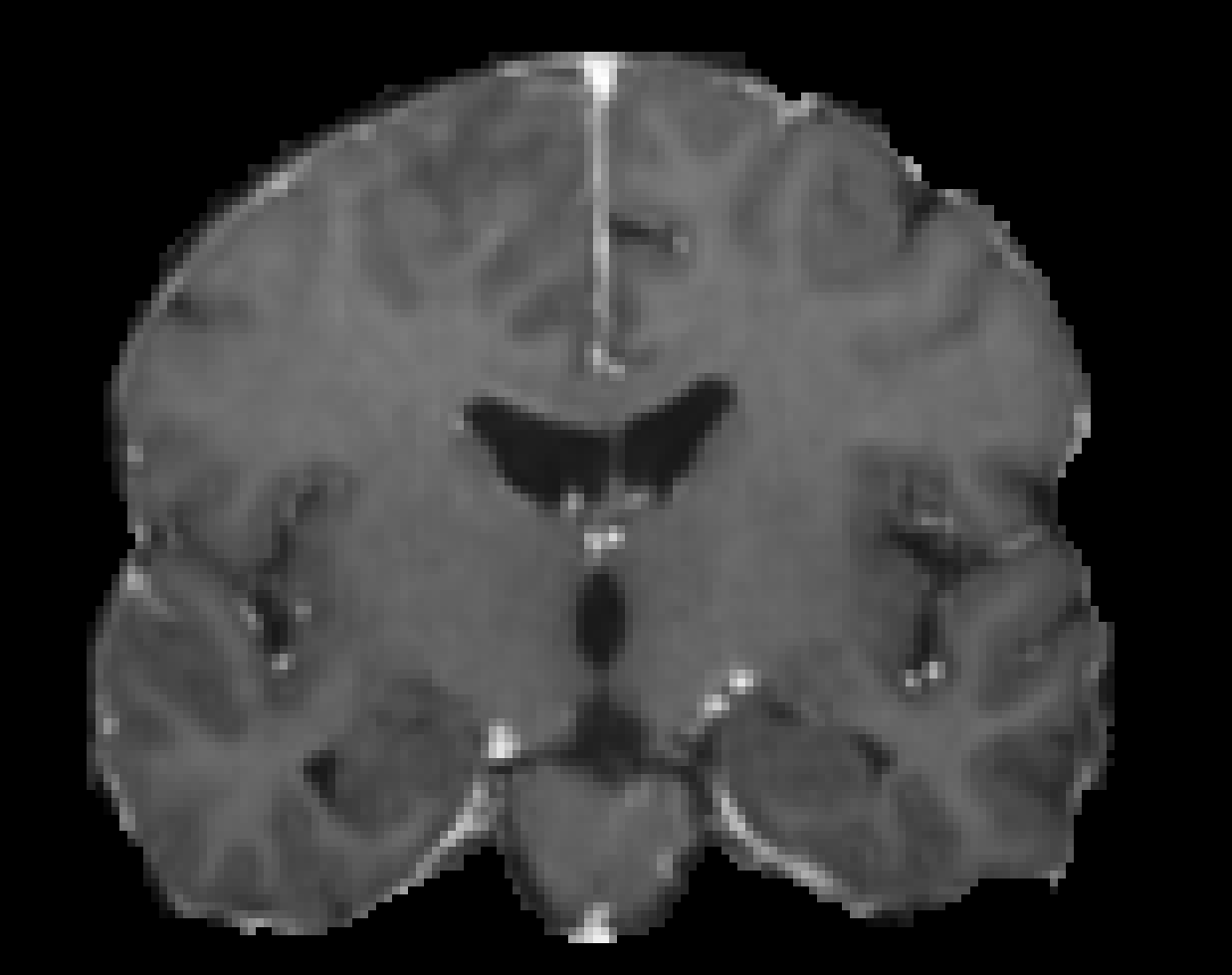}};
            \node[] at (4, 0)[anchor=south west]  {\includegraphics[width=2cm]{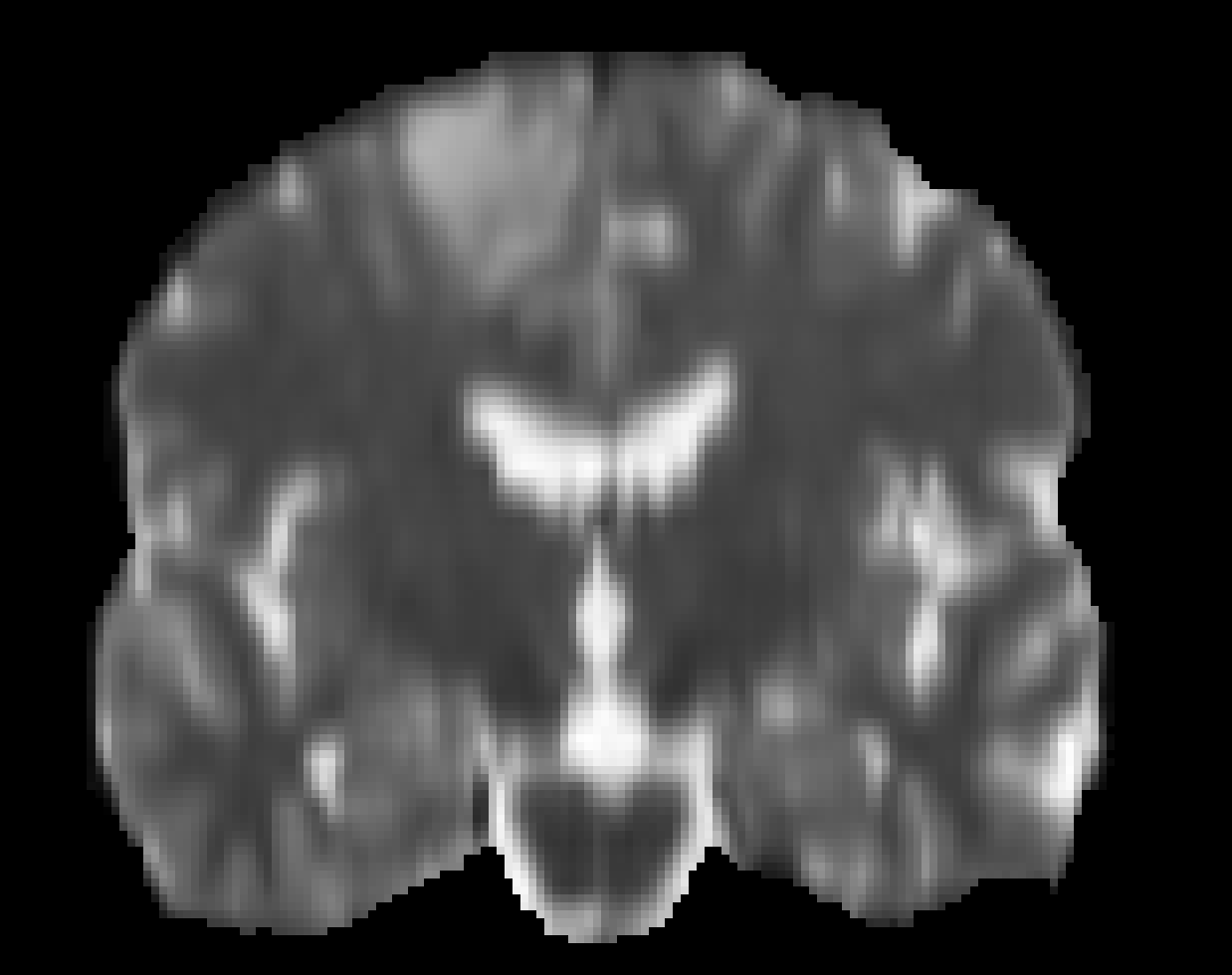}};
            \node[] at (6, 0)[anchor=south west]  {\includegraphics[width=2cm]{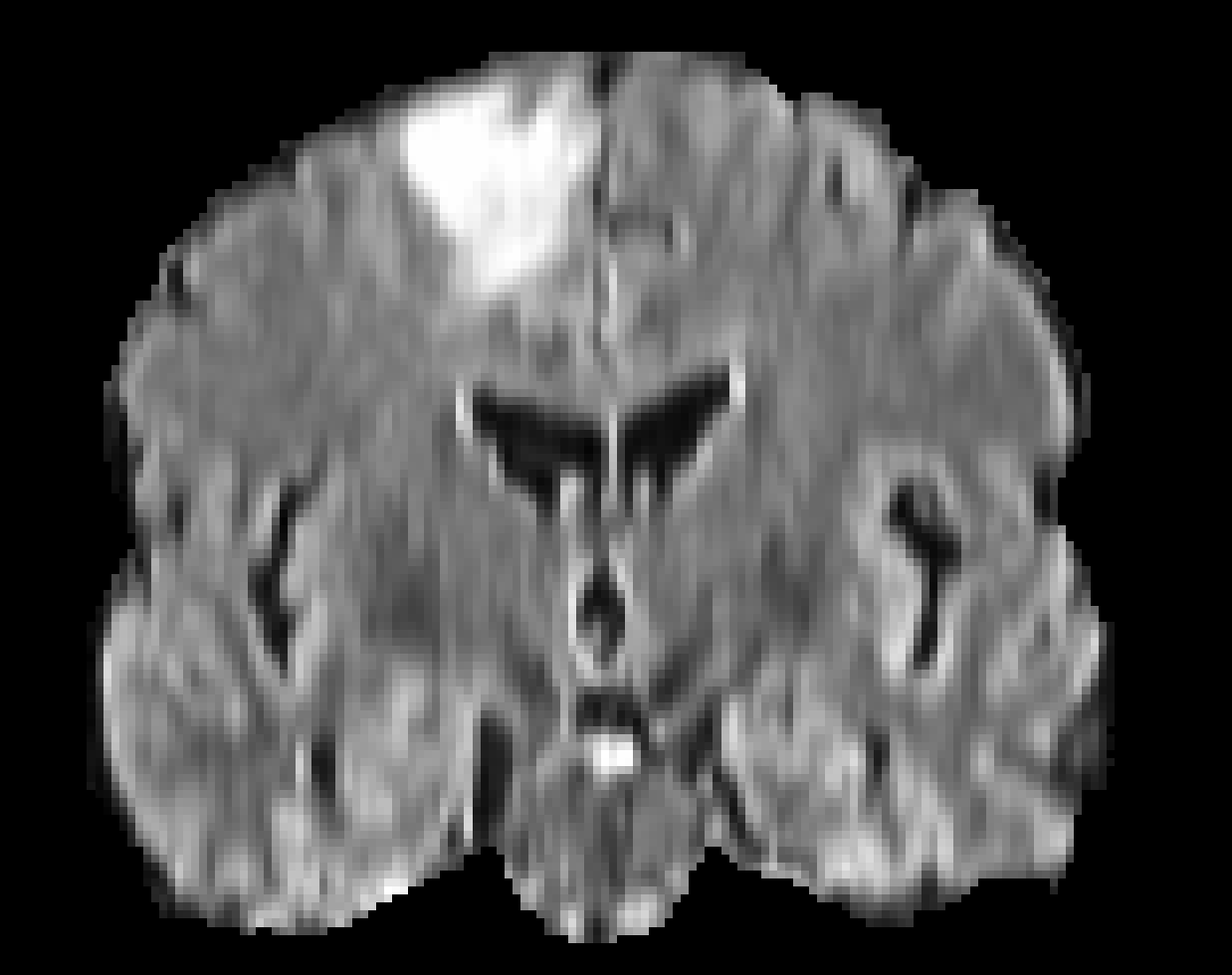}};
            \node[] at (0, 6)[anchor=south west]  {\includegraphics[width=2cm]{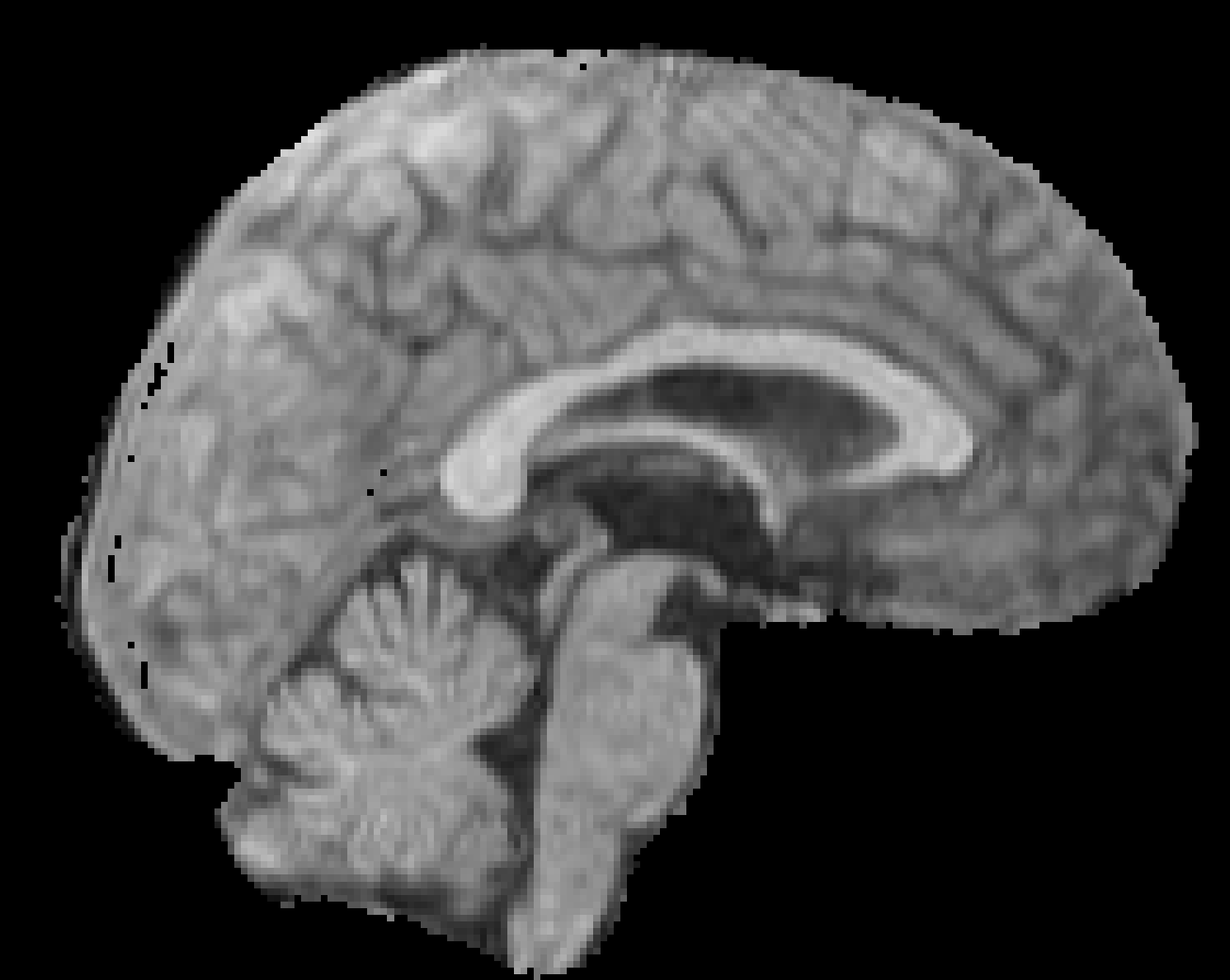}};
            \node[] at (2, 6)[anchor=south west]  {\includegraphics[width=2cm]{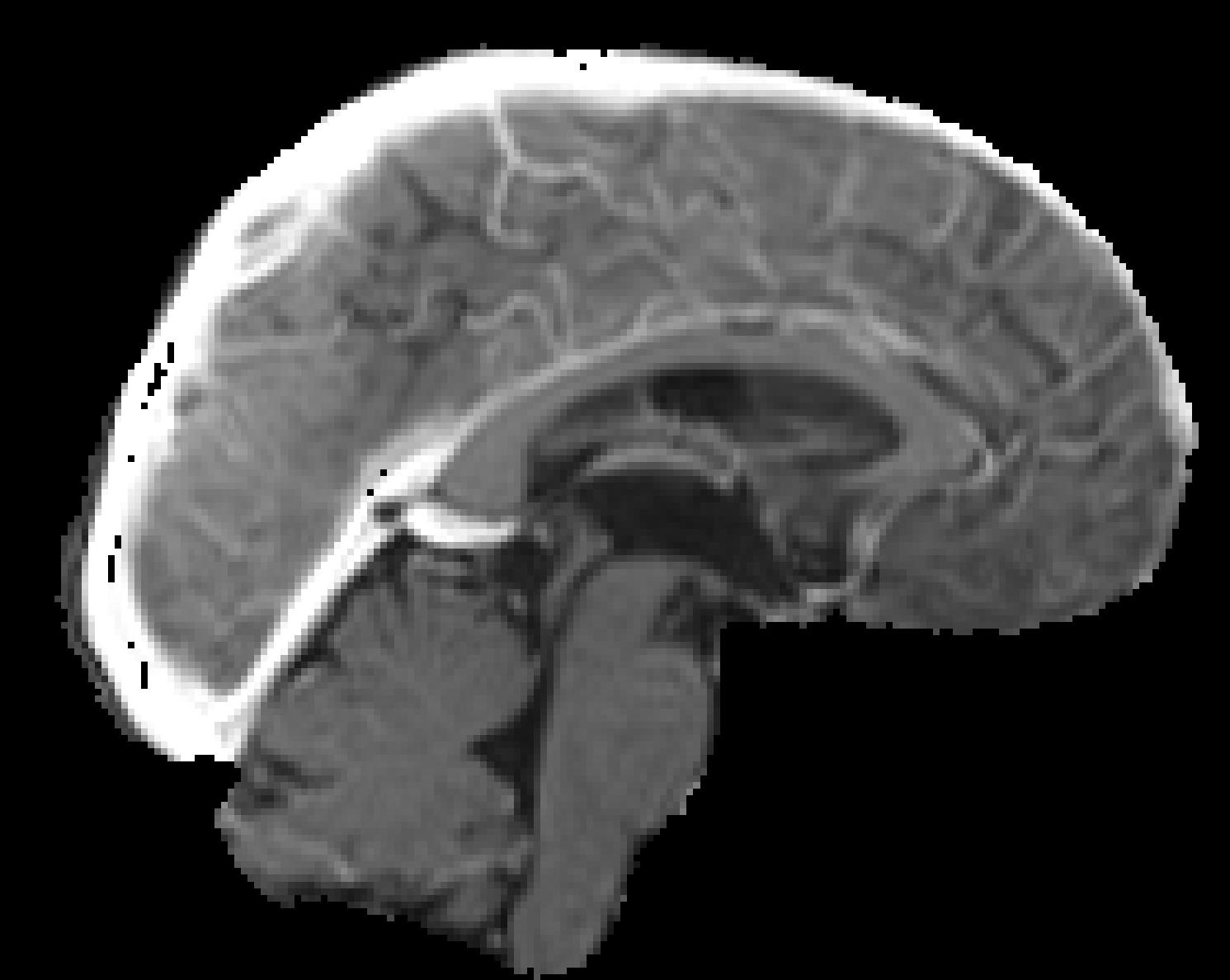}};
            \node[] at (4, 6)[anchor=south west]  {\includegraphics[width=2cm]{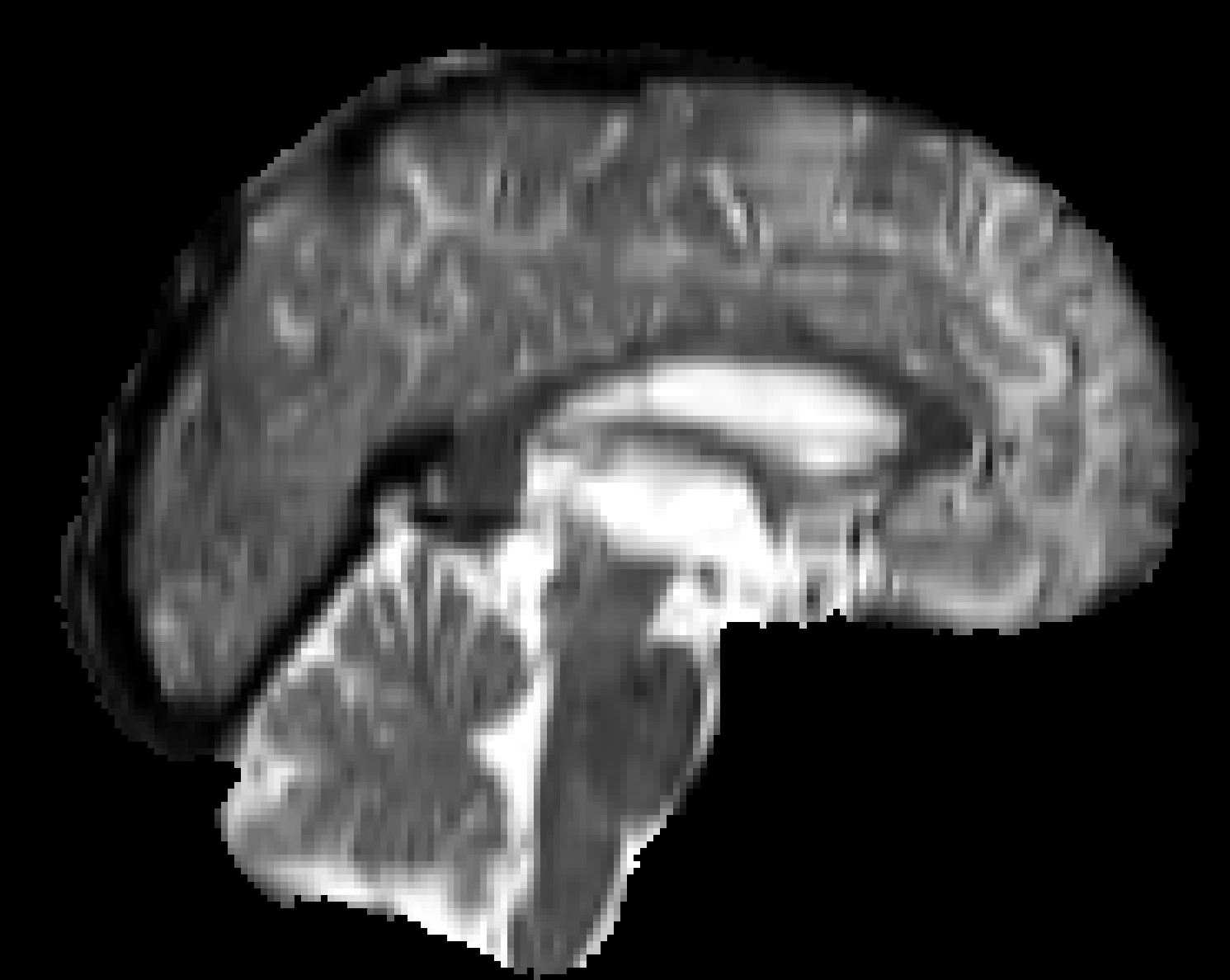}};
            \node[] at (6, 6)[anchor=south west]  {\includegraphics[width=2cm]{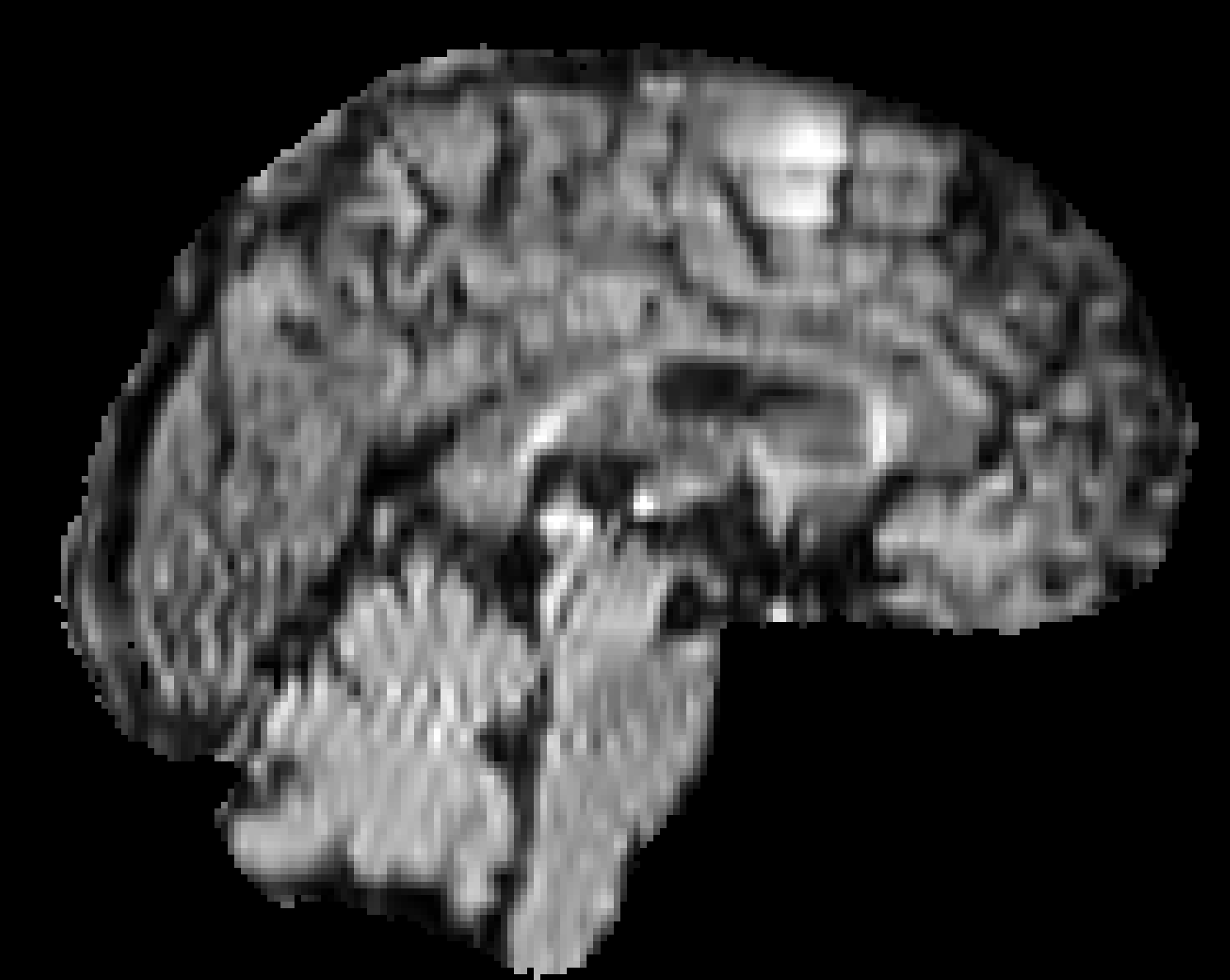}};
            \node[] at (0, 4)[anchor=south west]  {\includegraphics[width=2cm]{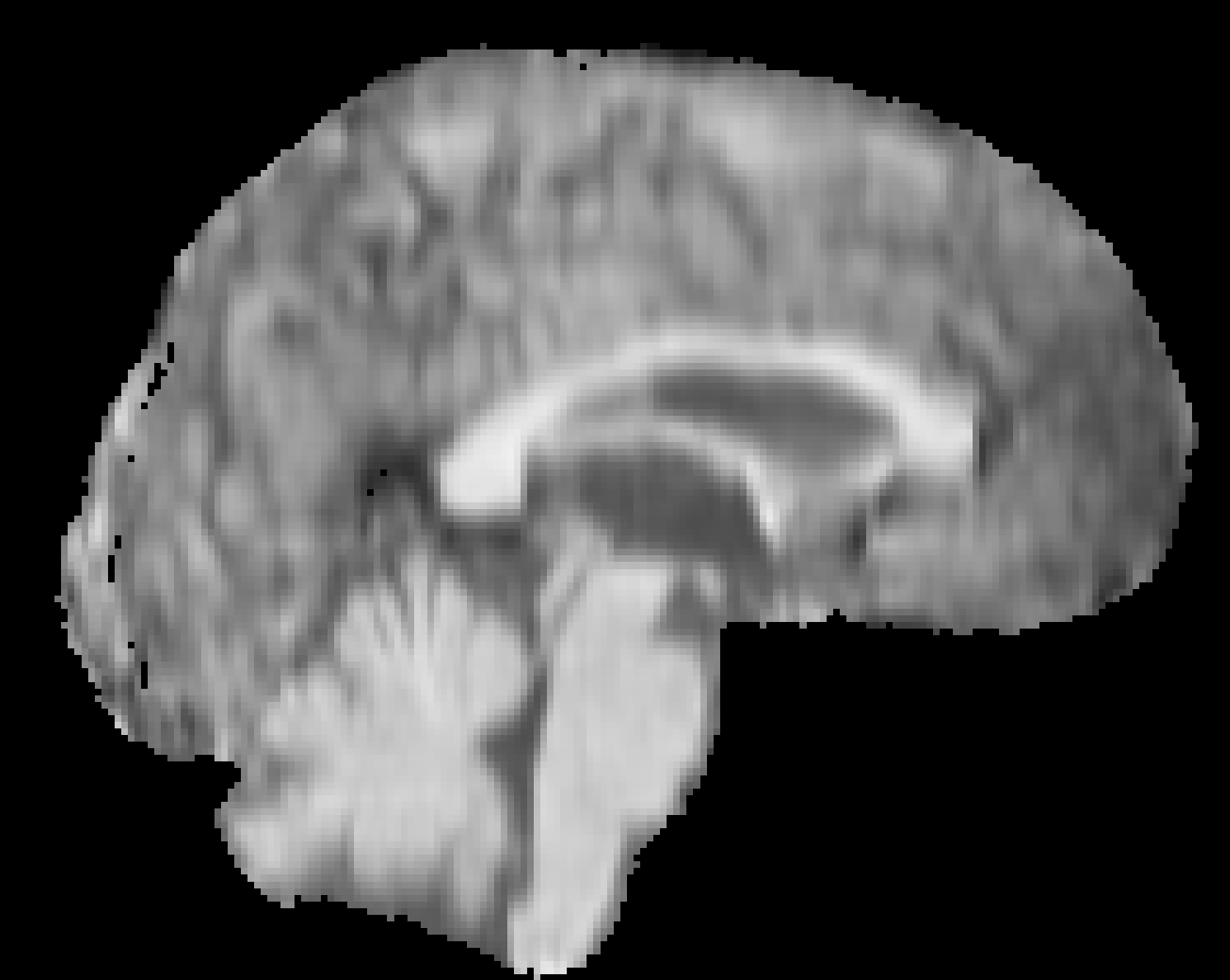}};
            \node[] at (2, 4)[anchor=south west]  {\includegraphics[width=2cm]{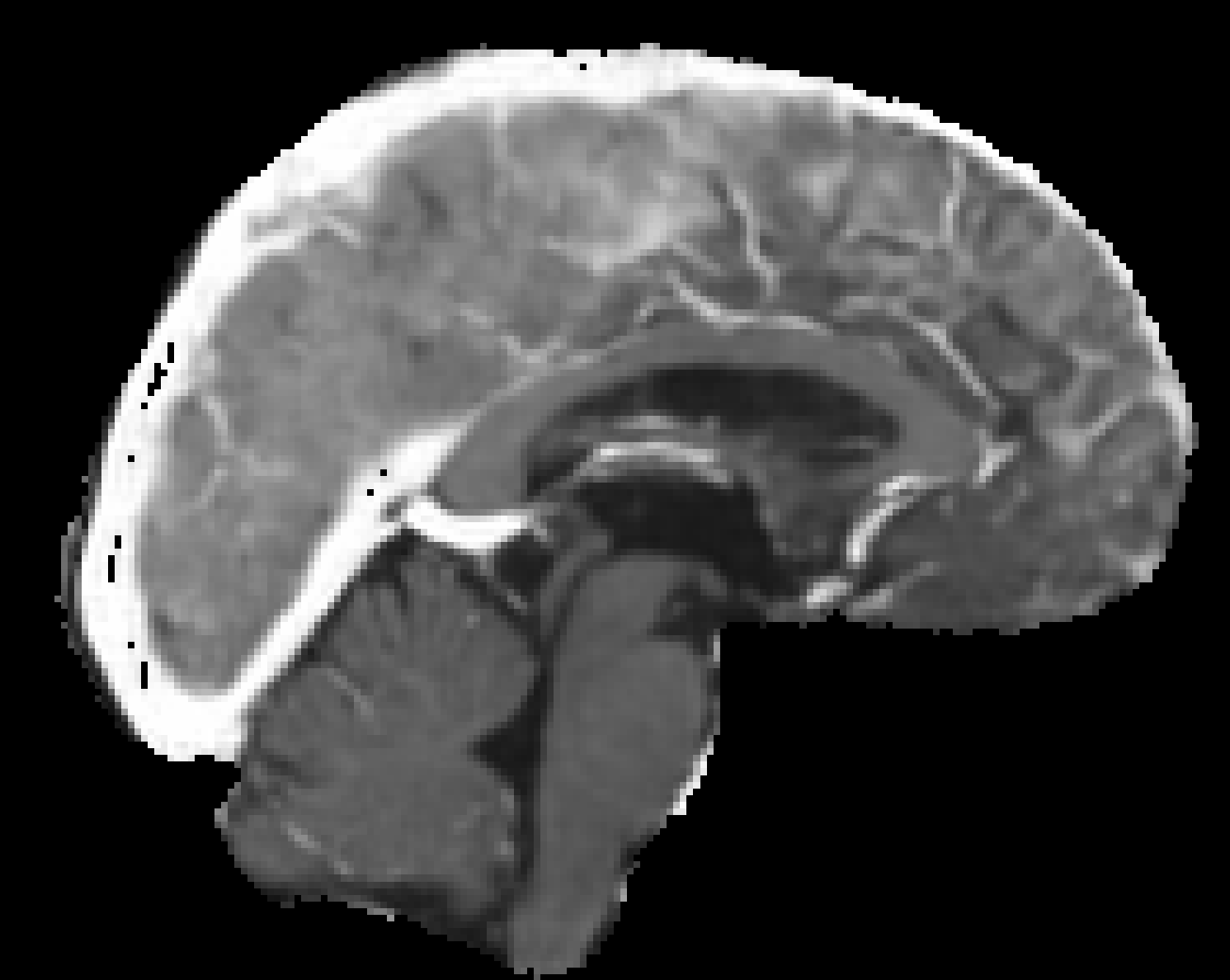}};
            \node[] at (4, 4)[anchor=south west]  {\includegraphics[width=2cm]{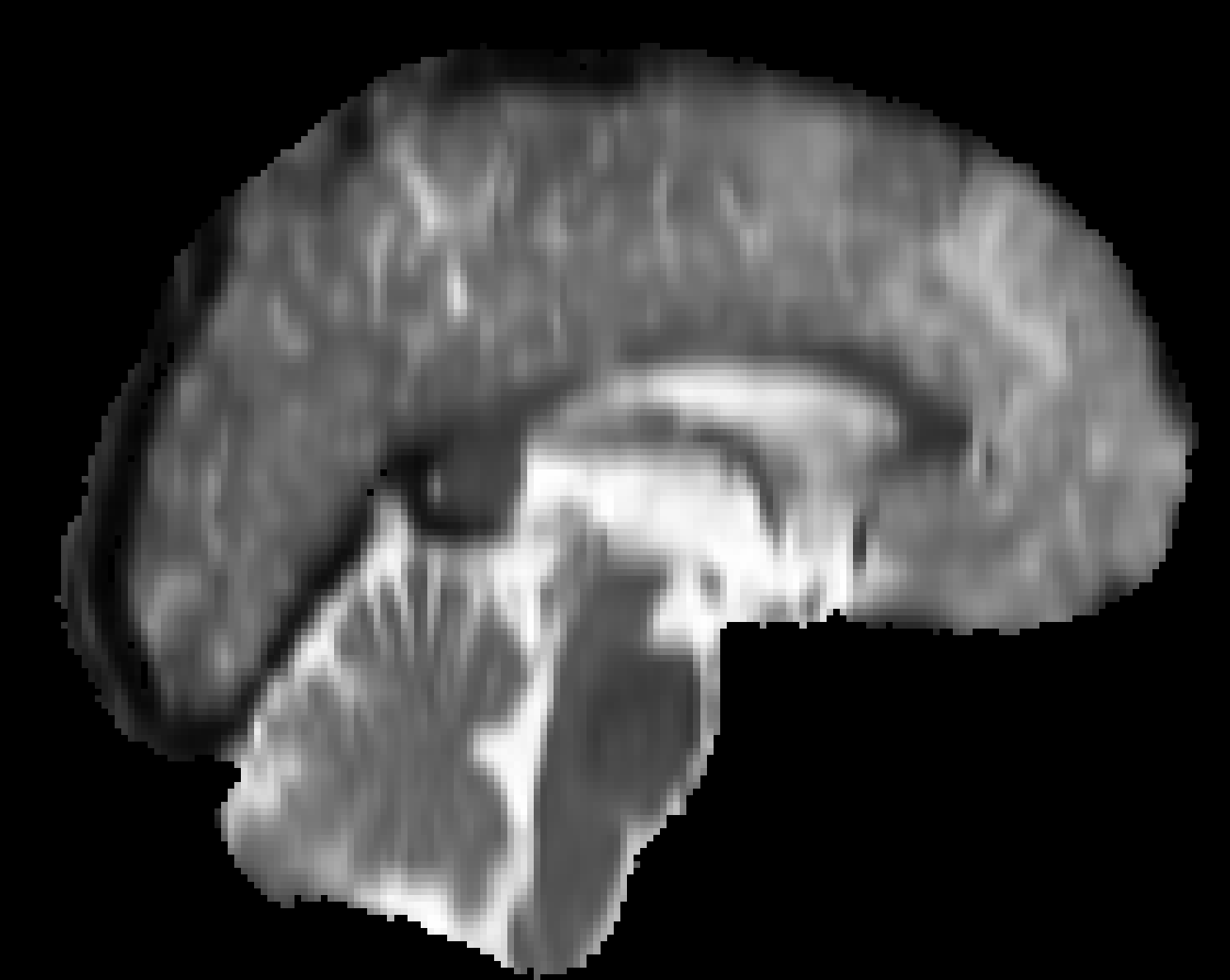}};
            \node[] at (6, 4)[anchor=south west]  {\includegraphics[width=2cm]{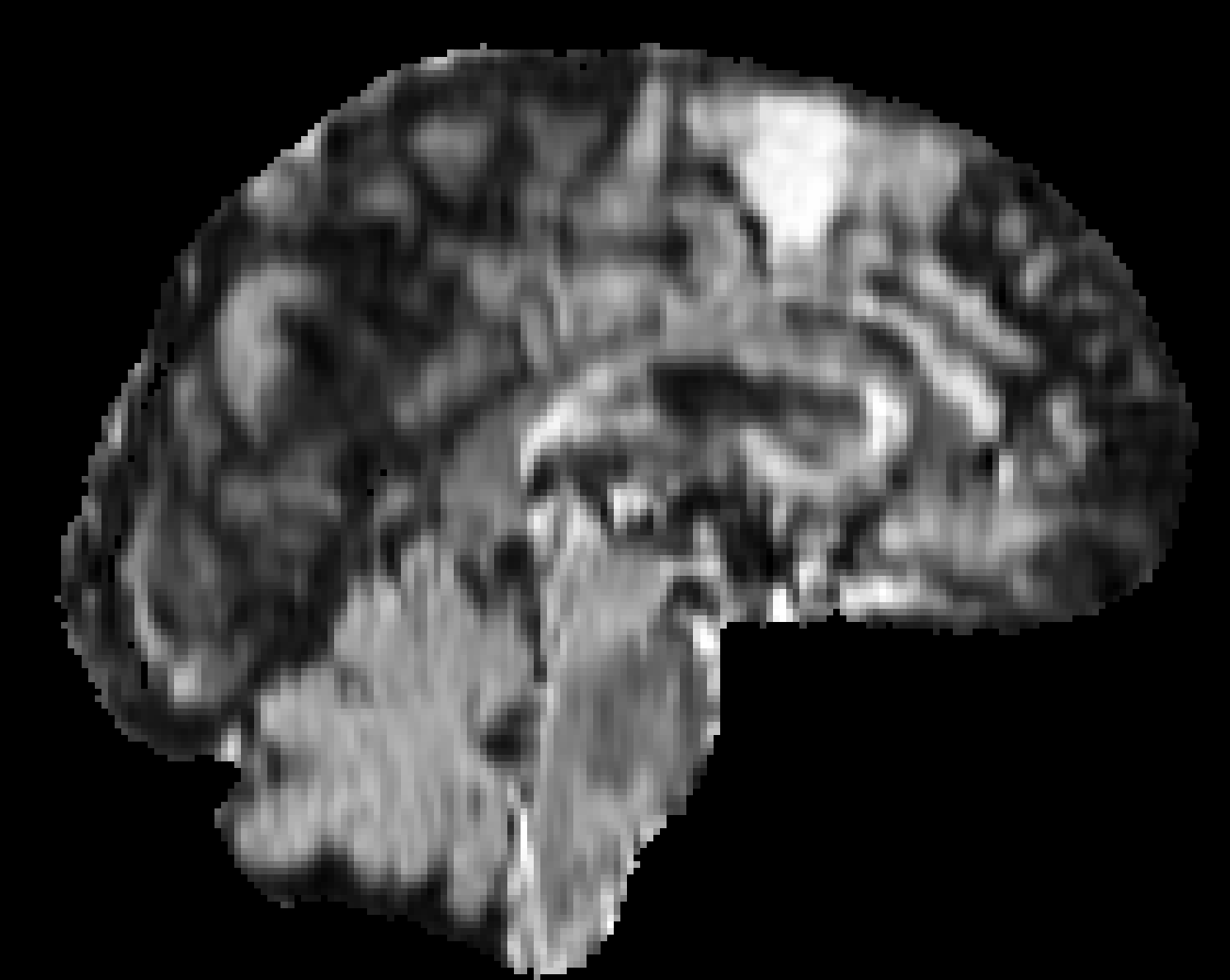}};
            \node[] at (0, 10)[anchor=south west]  {\includegraphics[width=2cm]{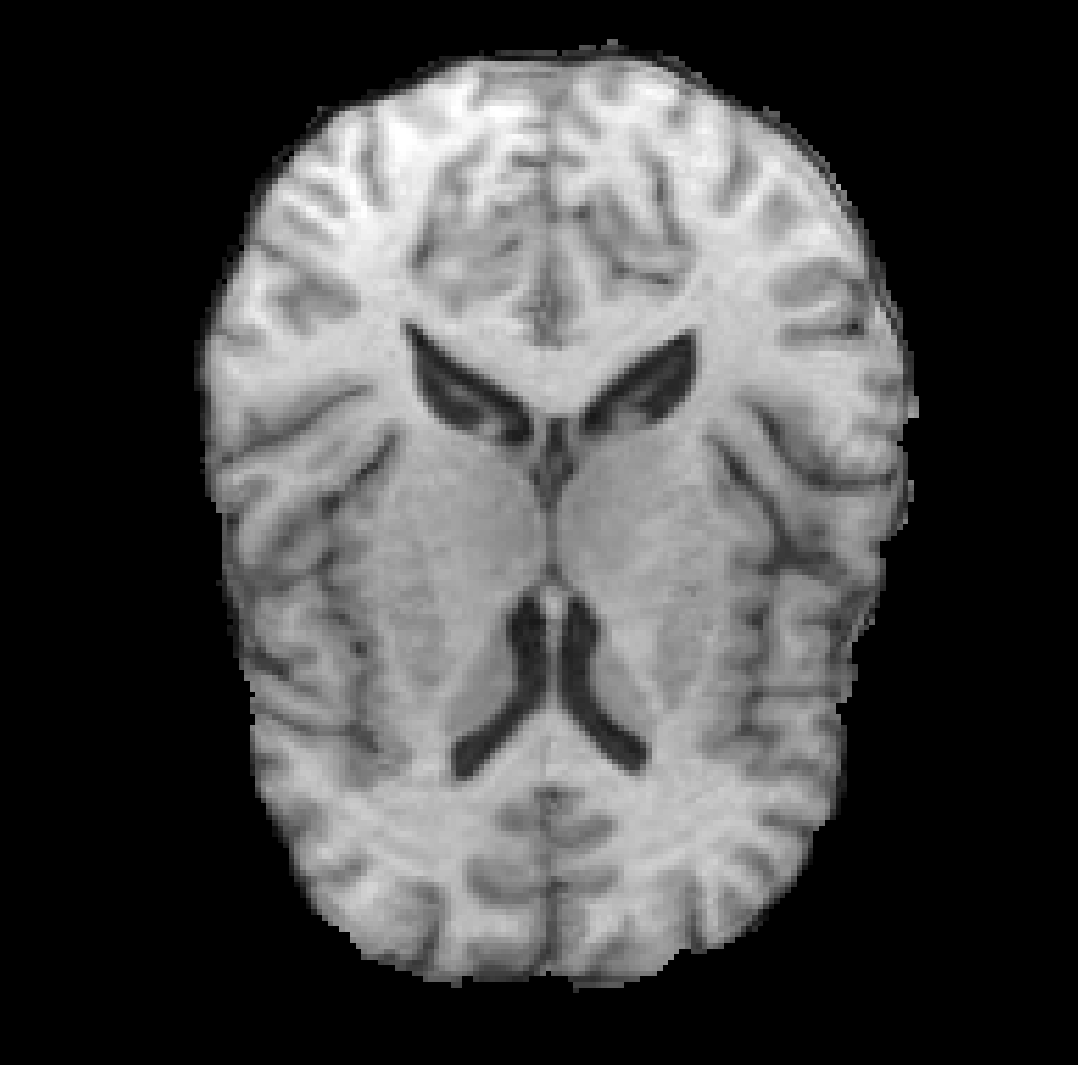}};
            \node[] at (2, 10)[anchor=south west]  {\includegraphics[width=2cm]{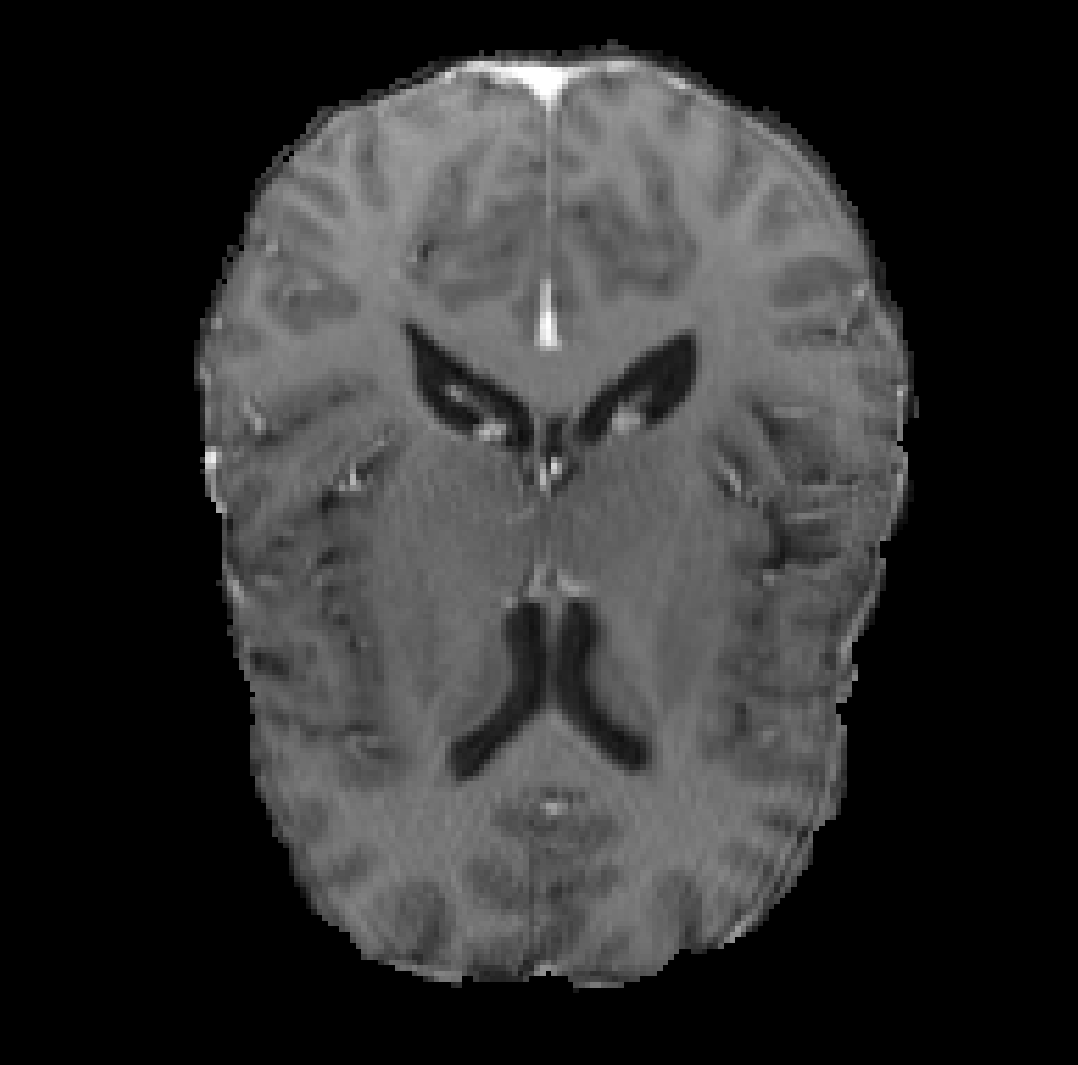}};
            \node[] at (4, 10)[anchor=south west]  {\includegraphics[width=2cm]{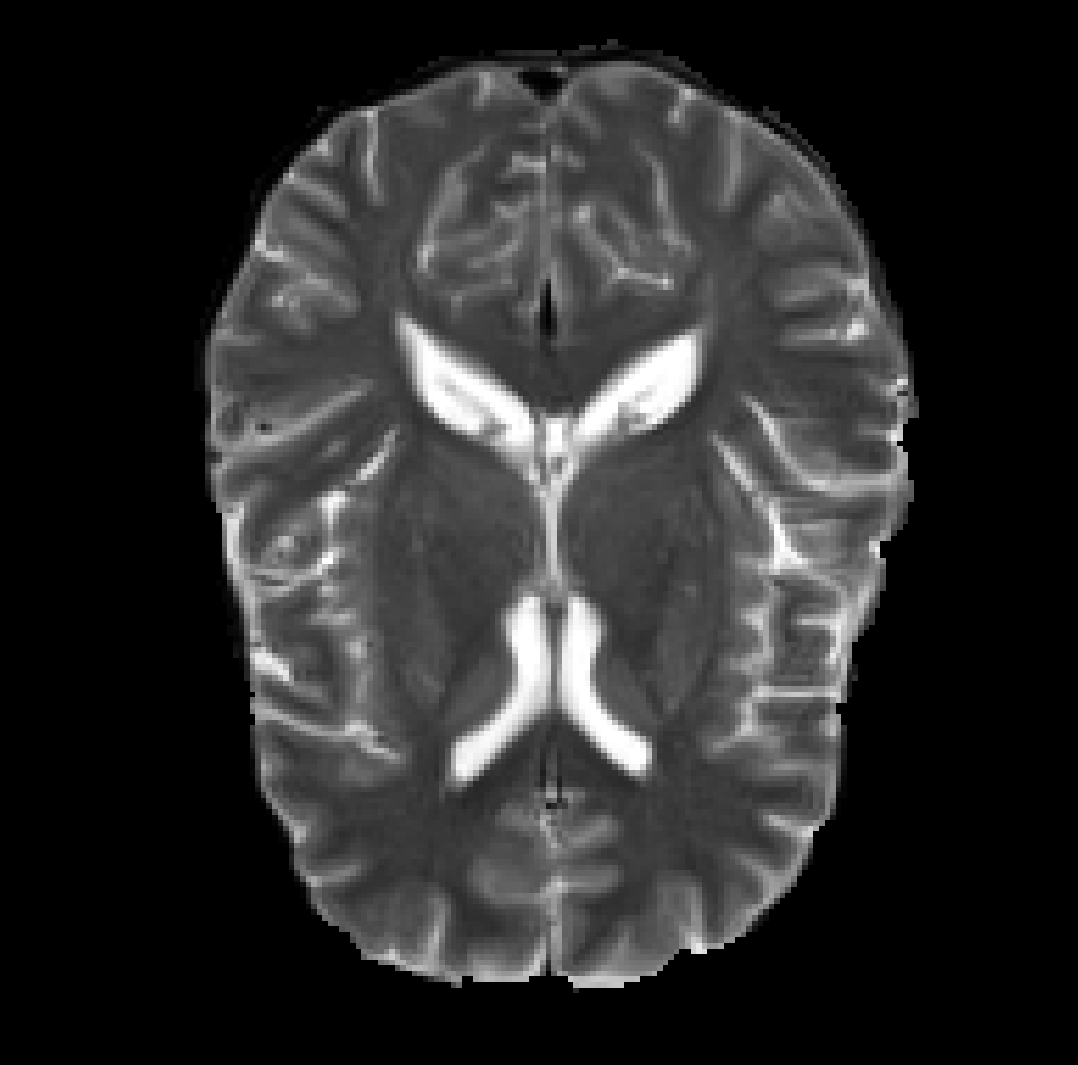}};
            \node[] at (6, 10)[anchor=south west]  {\includegraphics[width=2cm]{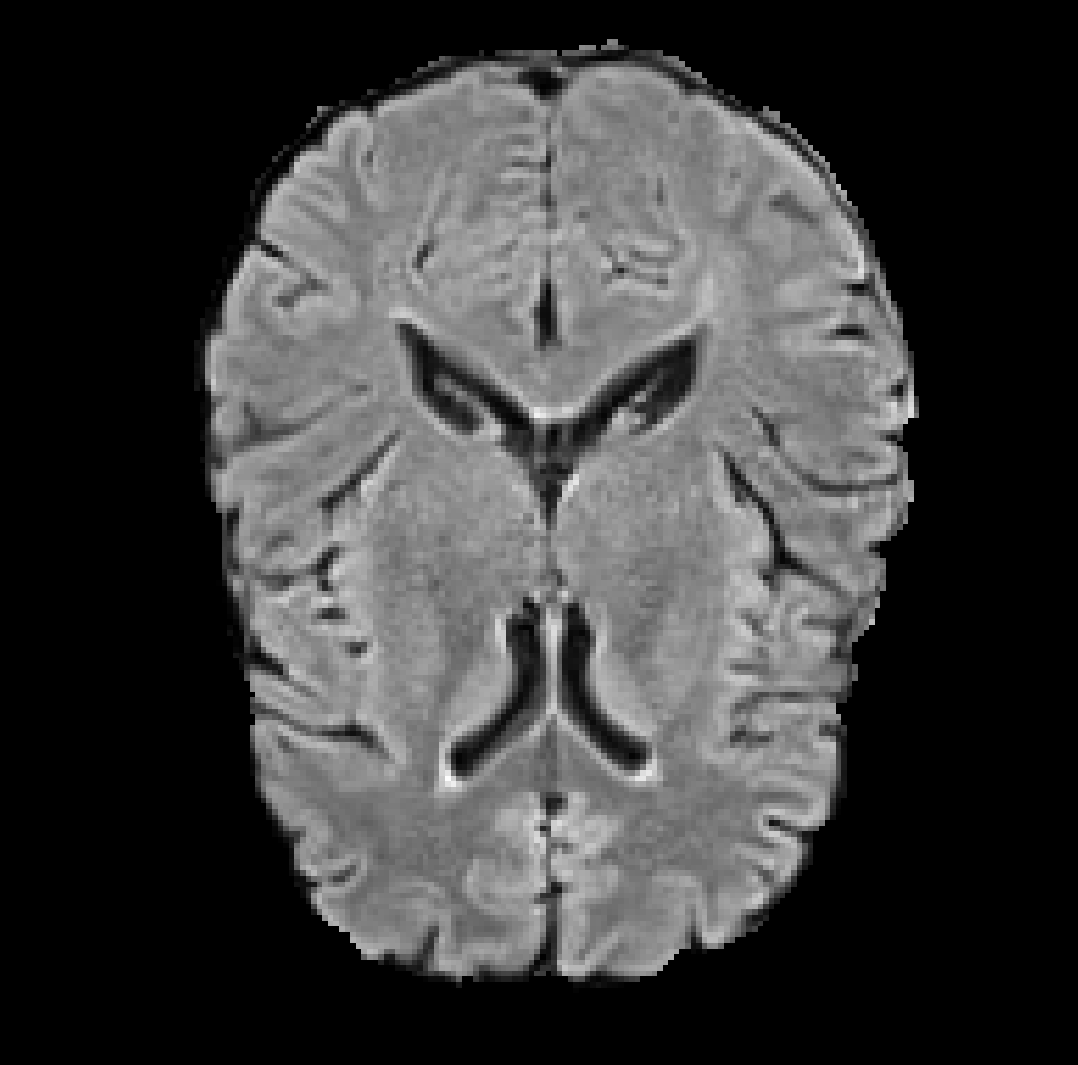}};
            \node[] at (0, 8)[anchor=south west]  {\includegraphics[width=2cm]{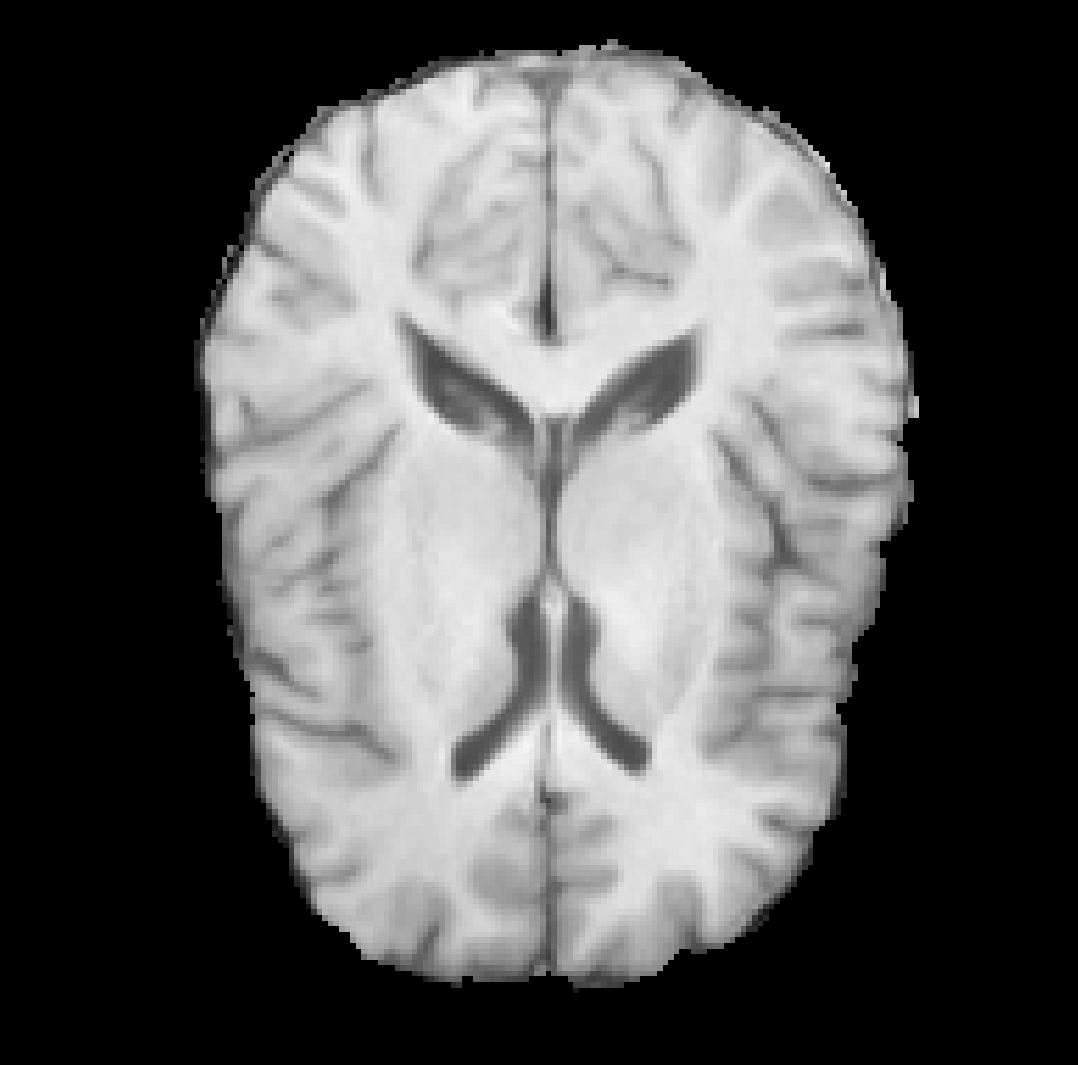}};
            \node[] at (2, 8)[anchor=south west]  {\includegraphics[width=2cm]{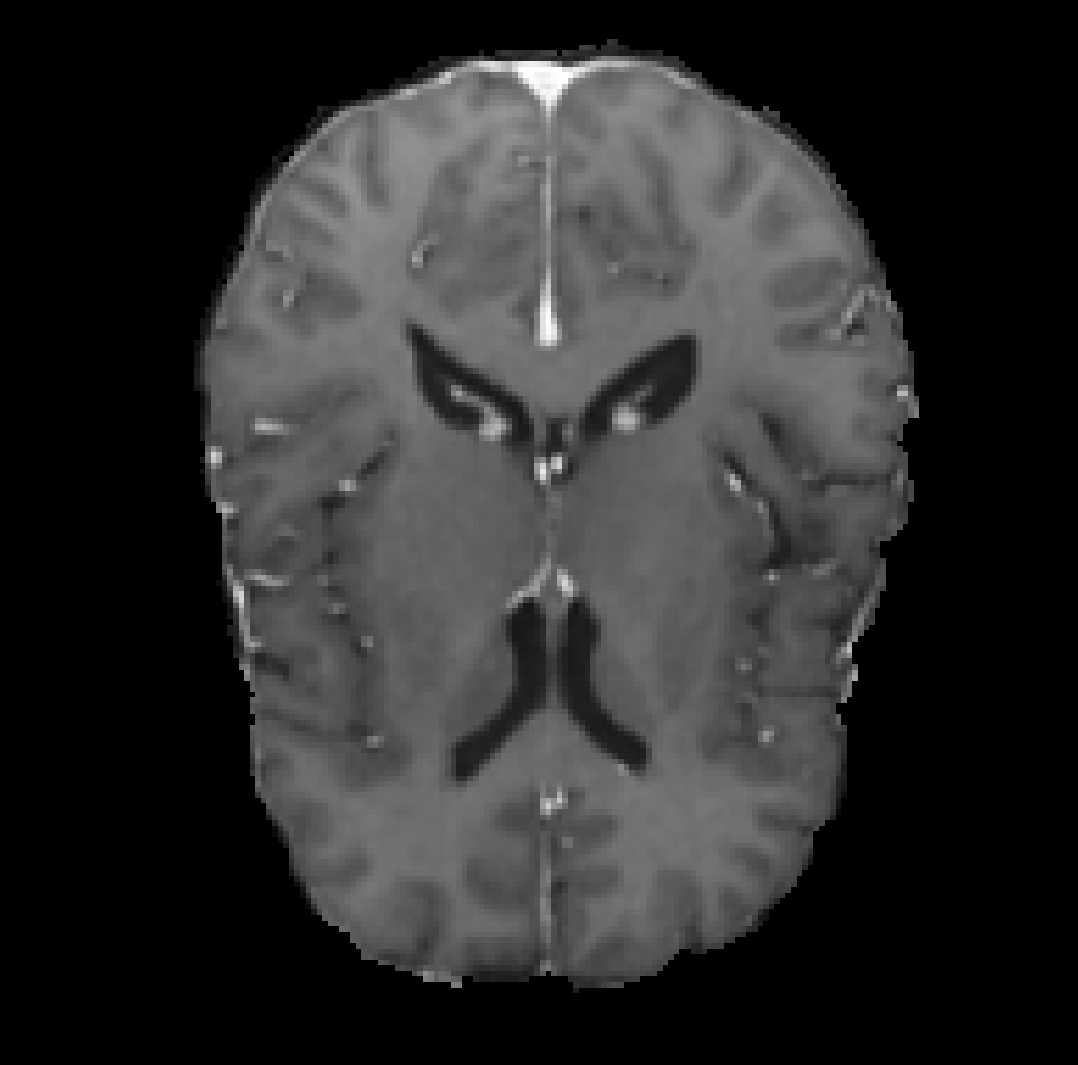}};
            \node[] at (4, 8)[anchor=south west]  {\includegraphics[width=2cm]{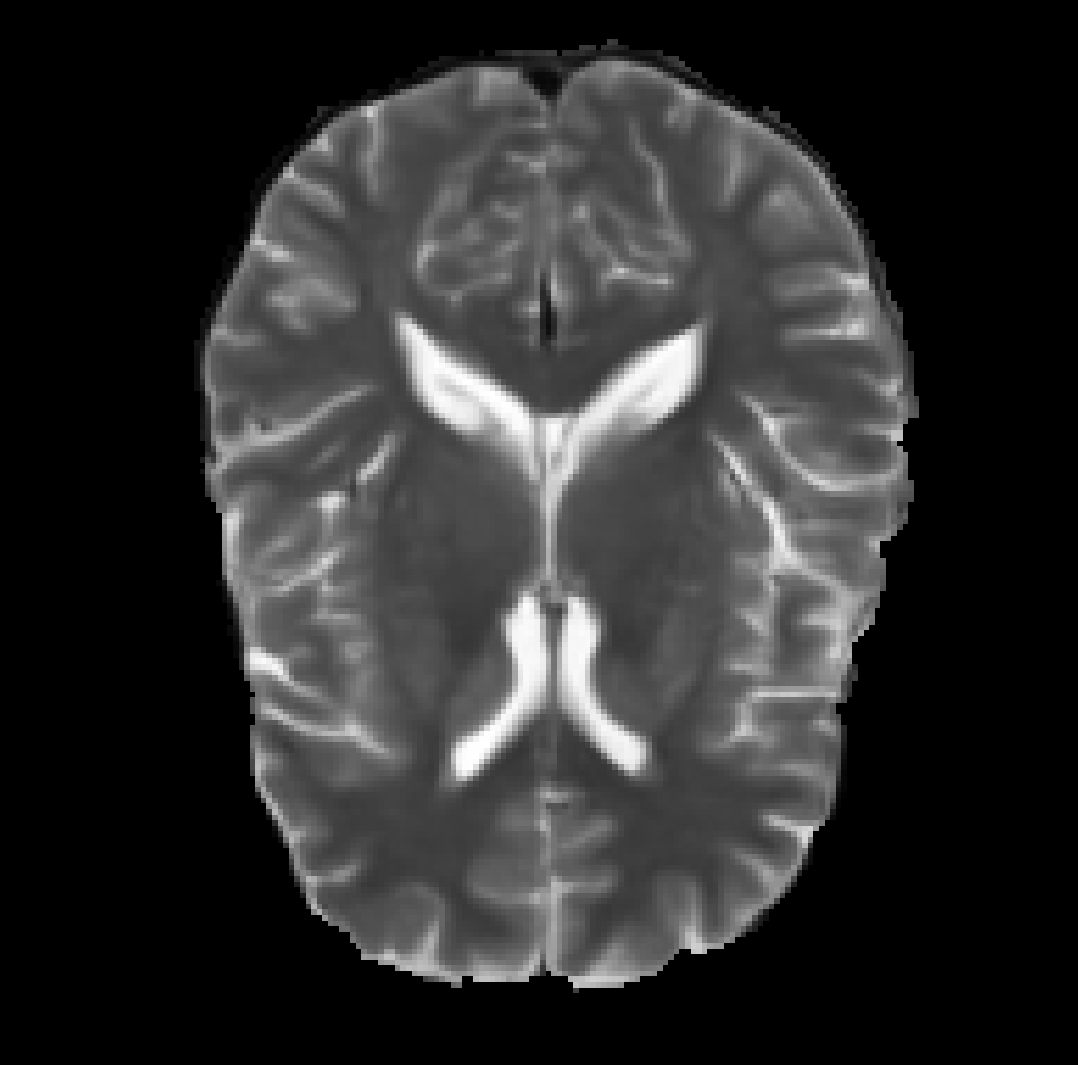}};
            \node[] at (6, 8)[anchor=south west]  {\includegraphics[width=2cm]{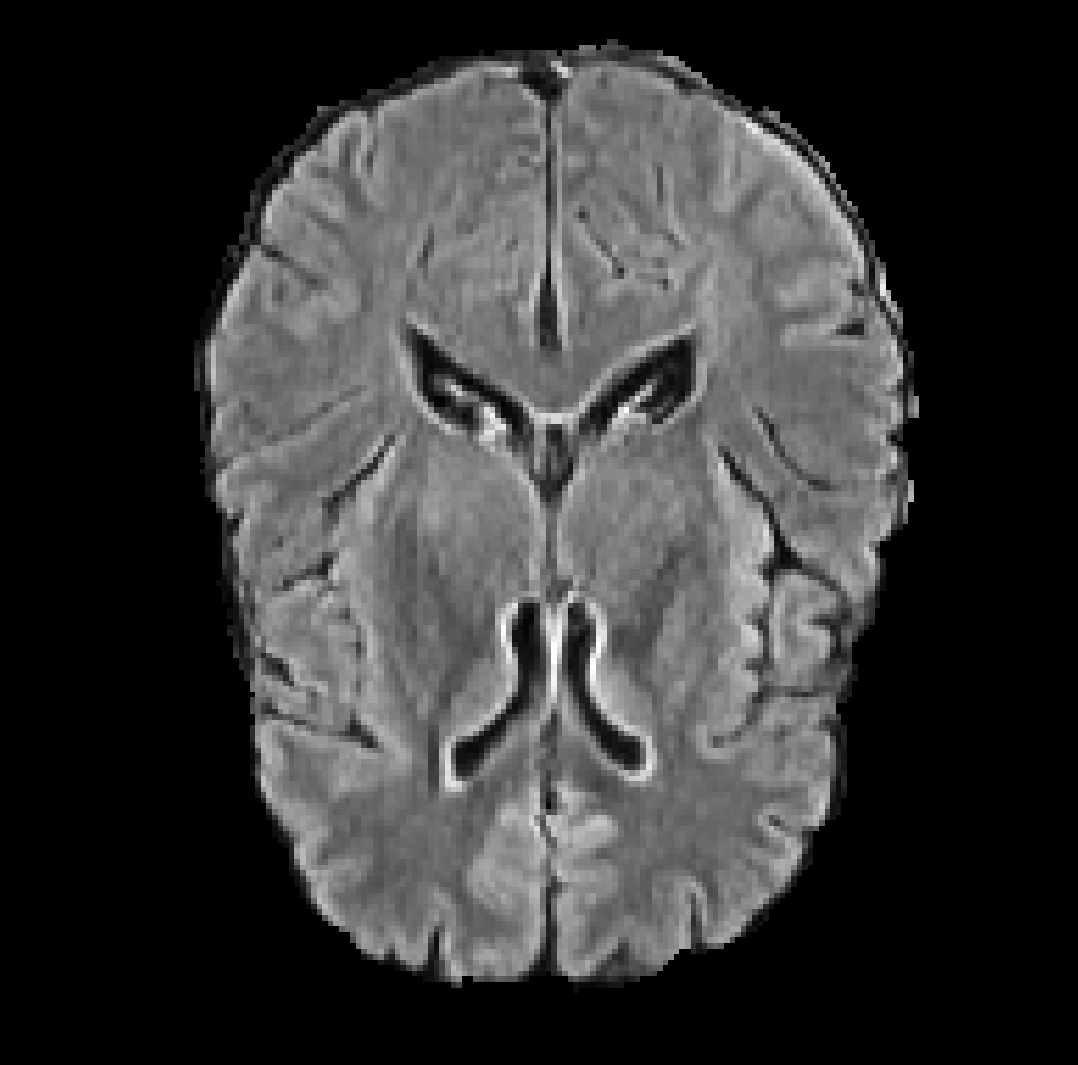}};
            \node[rotate=90] at (-0.2, 1)            {\scriptsize Synthetic};
            \node[rotate=90] at (-0.2, 3)            {\scriptsize Real};
            \node[rotate=90] at (-0.2, 5)            {\scriptsize Synthetic};
            \node[rotate=90] at (-0.2, 7)            {\scriptsize Real};
            \node[rotate=90] at (-0.2, 9)            {\scriptsize Synthetic};
            \node[rotate=90] at (-0.2, 11)            {\scriptsize Real};
            \node[] at (1,12.4) {\scriptsize T1};
            \node[] at (3,12.4) {\scriptsize T1ce};
            \node[] at (5,12.4) {\scriptsize T2};
            \node[] at (7,12.4) {\scriptsize FLAIR};
    	\end{tikzpicture}
     }
    \caption{Additional qualitative results of our proposed method. The synthetic images are generated conditioned on the real images from the three other modalities. We display the middle slice in the axial \emph{(top)}, sagittal \emph{(middle)}, and coronal \emph{(bottom)} plane.}
    \label{fig:results_2}
\end{figure}

Qualitative results of the synthesized images are shown in Fig.~\ref{fig:results_1} and Fig.~\ref{fig:results_2}, where we show all four synthesized contrasts. Each of the synthetic images was generated conditioned on the three given images.

\begin{table}
    \centering
    \caption{Image quality metrics for the four different models and the overall approach for randomly missing modalities.}
    \begin{tabular}{lccc}
        \textbf{Missing Modality} & MSE $\downarrow$ & PSNR $\uparrow$ & SSIM $\uparrow$\\
        \hline
        T1 & $1.65\times10^{-3}$ & $29.74$ & $0.956$\\
        T1ce &$2.85\times10^{-3}$ & $27.31$ & $0.936$\\
        T2 & $2.43\times10^{-3}$& $28.86$ & $0.952$\\
        FLAIR & $2.25\times10^{-3}$ & $27.83$ & $0.935$\\
        \hline
        Random & $1.82\times10^{-3}$ & $28.72$ & $0.946$\\
    \end{tabular}
    \label{tab:brats}
\end{table}

\subsection{Results on Test Data}
We will add quantitative evaluation scores (image quality and segmentation metrics) computed on the non-public test set containing 570 cases, as soon as they are provided by the challenge organizers.
\subsection{Ablation Study}
\label{subsec:ablation}
To find a model that fits the image-to-image translation task, we ablate different setups by varying the type of skip connection used in the denoising network $\epsilon_{\theta}$, the number of base channels $C$, as well as the applied variance schedule. The results of this ablation study are shown in Tab.~\ref{tab:ablation}. We empirically found that a setup with standard skip connections with concatenation, a linear variance schedule and $C=64$ base channels performed best.

\begin{table}
    \centering
    \caption{Ablation study of different model setups on the T1-weighted brain MR generation task. We compare different skip connections, variance schedules, and different numbers of base channels. We measure MSE, PSNR, SSIM, inference time and inference GPU memory footprint. \textbf{Bold} is best, \underline{underline} is second best, \colorbox{lightgray}{gray} is the overall best performing setup. The scores were computed on slightly cropped images with a resolution of $155 \times 224 \times 224$, to reduce the influence of black background voxels.}
    \resizebox{\textwidth}{!}{
    \begin{tabular}{llc|ccccc}
        Skip Connection & $\beta$-Schedule & Base Channels $C$ & MSE ($\downarrow$) & PSNR ($\uparrow$) & SSIM ($\uparrow$) & Time [s] & Memory [MB]\\\hline
        addition & linear & 128 & $1.73\times10^{-3}$ & $28.99$ & $\underline{0.948}$ & $\underline{283}$ & $\textbf{8262}$ \\
        addition & cosine & 128 & $1.95\times10^{-3}$ & $28.31$ & $0.942$ & $\underline{283}$ & $\textbf{8262}$ \\
        \rowcolor{lightgray} concatenation & linear & 64 & $\underline{1.65\times10^{-3}}$ & $\textbf{29.14}$ & $\textbf{0.949}$ & $\textbf{172}$ & $\underline{9578}$\\
        concatenation & linear & 96 & $1.72\times10^{-3}$ & $\underline{29.06}$ & $0.947$ & $320$ & $13434$\\
        concatenation & cosine & 64 & $\mathbf{1.62\times10^{-3}}$ & $\underline{29.06}$ & $\underline{0.948}$ & $\textbf{172}$ & $\underline{9578}$\\
        concatenation & cosine & 96 & $1.77\times10^{-3}$ & $28.73$ & $0.945$ & $320$ & $13434$ \\
    \end{tabular}
    }
    \label{tab:ablation}
\end{table}
\section{Conclusion}
In this paper, we introduce \textbf{cWDM}, a conditional Wavelet Diffusion Model for cross-modality image synthesis. We adopted the Wavelet Diffusion Model for high-resolution medical image synthesis, proposed in \cite{friedrich2024wdm}, to tackle the paired image-to-image translation task on full-resolution volumes. Our qualitative and quantitative results suggest that our method effectively addresses the issue of missing MR images and enables the application of well-performing segmentation models in clinical settings. In addition, the presented approach could be applied to other paired image-to-image translation tasks, such as CT~$\leftrightarrow$~MR and MR~$\leftrightarrow$~PET translation, or mask-conditioned anatomically guided image generation. 
\begin{credits}
\subsubsection{\ackname} This work was financially supported by the Werner Siemens Foundation through the MIRACLE II project.
\subsubsection{\discintname}
The authors have no competing interests to declare that are relevant to the content of this article.
\end{credits}
\newpage
\bibliographystyle{splncs04.bst}
\bibliography{bibliography.bib}
\end{document}